\newif\ifShowKeys
\ShowKeystrue

\documentclass[letterpaper,10pt]{article}

\pdfoutput=1
\usepackage[no-natbib-sort]{my-jheppub}

\usepackage{amsmath, amssymb}

\usepackage{amsthm}

\usepackage{bm,environ,mathrsfs,array,arydshln,booktabs,float,slashed,hyperref}
\usepackage[mathcal]{euscript}
\usepackage{tensor} 	

\usepackage[nodayofweek]{date time}

\usepackage{graphicx,epsfig}
\usepackage{epic}
\usepackage{tikz} 
\usetikzlibrary{decorations.pathreplacing,decorations.markings,snakes}

\tikzset{middlearrow/.style={
        decoration={markings,
            mark= at position 0.5 with {\arrow{#1}} ,
        },
        postaction={decorate}
    }
}

\usepackage{floatflt}

\usepackage{aurical}
\usepackage[T1]{fontenc}

\usepackage{framed}
\definecolor{shadecolor}{RGB}{255, 230, 204}

\allowdisplaybreaks

\newcommand{\red}[1]{\textcolor{red}{#1}}

\newcommand{\be}{\begin{equation}}
\newcommand{\ee}{\end{equation}}
\newcommand{\mc}{\mathcal }

\newcommand{\la}{\label}
\newcommand{\eps}{\varepsilon}

\newcommand{\de}{\Delta}

\makeatletter
\DeclareFontFamily{OMX}{MnSymbolE}{}
\DeclareSymbolFont{MnLargeSymbols}{OMX}{MnSymbolE}{m}{n}
\SetSymbolFont{MnLargeSymbols}{bold}{OMX}{MnSymbolE}{b}{n}
\DeclareFontShape{OMX}{MnSymbolE}{m}{n}{
    <-6>  MnSymbolE5
   <6-7>  MnSymbolE6
   <7-8>  MnSymbolE7
   <8-9>  MnSymbolE8
   <9-10> MnSymbolE9
  <10-12> MnSymbolE10
  <12->   MnSymbolE12
}{}
\DeclareFontShape{OMX}{MnSymbolE}{b}{n}{
    <-6>  MnSymbolE-Bold5
   <6-7>  MnSymbolE-Bold6
   <7-8>  MnSymbolE-Bold7
   <8-9>  MnSymbolE-Bold8
   <9-10> MnSymbolE-Bold9
  <10-12> MnSymbolE-Bold10
  <12->   MnSymbolE-Bold12
}{}

\let\llangle\@undefined
\let\rrangle\@undefined
\DeclareMathDelimiter{\llangle}{\mathopen}%
                     {MnLargeSymbols}{'164}{MnLargeSymbols}{'164}
\DeclareMathDelimiter{\rrangle}{\mathclose}%
                     {MnLargeSymbols}{'171}{MnLargeSymbols}{'171}
\makeatother

\title{On  boundary correlators  in  Liouville theory on  AdS$_{2}$    
}

\author[a]{Matteo Beccaria,} 
\author[b]{ and \ \ Arkady A. Tseytlin\footnote{Also at the   Institute  for Theoretical and Mathematical Physics, Moscow State University
and  Lebedev Institute, Moscow.}} 
\abstract{
We consider  the  Liouville  theory in fixed  Euclidean AdS$_2$   background.
Expanded  near the  minimum of the 
potential the elementary field  has mass squared 2 
 and  (assuming the standard Dirichlet b.c.)  corresponds  to a  dimension 2   operator at the boundary.
 We provide strong evidence for the  conjecture that the boundary correlators 
 of the Liouville field  are the same as the correlators of the   holomorphic stress  tensor 
 (or the  Virasoro  generator  with the same
 central charge)
 on a half-plane or a disc   restricted to the  boundary. 
 This relation was first  observed 
  at the leading semiclassical order (tree-level Witten diagrams in AdS$_2$)
  in arXiv:1902.10536 
 and  here we demonstrate its validity also  at the one-loop level. 
 We also  discuss    arguments that may lead to its general proof.
}

\affiliation[a]{Dipartimento di Matematica e Fisica Ennio De Giorgi,\\
Universit\`a del Salento \& INFN, Via Arnesano, 73100 Lecce, 
Italy} 
\affiliation[b]{Blackett Laboratory, Imperial College, London SW7 2AZ, U.K.}
\emailAdd{matteo.beccaria@le.infn.it, \ tseytlin@imperial.ac.uk}

\begin{document}

\date{\currenttime}

\begin{tabbing}
\hspace*{11.7cm} \=  \kill 
    \> Imperial-TP-AT-2019-02  
\end{tabbing}

\maketitle


 \def\ads  {AdS$_2$ }
\def \ov {\over}\def \te  {\textstyle} 
\def \ci {\cite}
\def \foot {\footnote}
\def \b{\beta}
\def \m {\mu}
\def \n {\nu}
\def \del{\partial}
\def \p {\phi}
\def \ep{\epsilon}
\newcommand{\rf}[1]{(\ref{#1})}
\def \r {\rho}
\def \k {\kappa}
\def \l {\lambda}
\def\z{\zeta}
\def \iffa {\iffalse}
\def \d {\partial} \def \no {\notag}\def \OO {{\cal{O}}}
\def \a  {\alpha} \def \ha {{\textstyle{1 \ov 2}}}
\def \sql {\sqrt{\lambda}}
\def \ed {
\bibliography{BT-Biblio}
\bibliographystyle{JHEP}
\end{document}}
\def \bz {\mathsf{z}}\def \vp {\varphi}
\def \adsf {AdS$_2$/CFT$_1$  } \def \P {\Phi}
\def \rr  {{\rm q}}
\def \widetildeB  {{\rm B}}
\def  \ZZ {{\sf ZZ\ }} 
\def \bb {{\rm b}}
 \def \C {{\rm q}}
\def \vt  {\vartheta} \def \vtheta {\vartheta}
\def \rC {{\rm q}}
\def \B {\Sigma} \def \rB {{\rm B}}
\def \CC  {{\rm C}}
\def \RR  {\mathbb R}


\section{Introduction} 

Study of quantum field theories in  AdS$_{2}$   background is of interest   from several points of view
(see, e.g.,  \ci{DHoker:1983zwg,DHoker:1983msr,Inami:1985di,Callan:1989em,Zamolodchikov:2001ah,Carmi:2018qzm}
and references  there). 
A recent example  appeared in  the   investigation of correlators
of operators inserted  on a  straight or circular  Wilson line
in the context of  AdS$_5$/CFT$_4$  correspondence
 \cite{Drukker:2000ep,Alday:2007he,Polyakov:2000ti,Polchinski:2011im, 
Drukker:2006xg,Giombi:2017cqn,Beccaria:2018ocq,Beccaria:2019dws}. As discussed in \ci{Giombi:2017cqn}, 
starting with the string action in  AdS$_{5}\times S^5$  and  expanding it  near the corresponding  minimal surface 
one  gets a 2d field theory action in  AdS$_2$ background. The  boundary correlators  of elementary excitations 
that should   match the  strong coupling   limit of  the  Wilson loop "defect" correlators in $\mathcal N=4$ super Yang-Mills theory
  are constrained  by the isometry of AdS$_2$ or the 1d conformal group $SO(2,1)\simeq SL(2,\RR)$  providing a non-trivial 
 example  of a  non-gravitational     AdS$_2$/CFT$_1$ duality.
One  technical problem in the approach of  \ci{Giombi:2017cqn,Beccaria:2019dws}
 is how to systematically compute   loop corrections to boundary correlators  of 2d quantum  fields in AdS$_2$.

Using this as a partial motivation,  
 it  is  interesting to   study    loop corrections to 
boundary  correlators   in  some  simple  2d QFTs  defined  on  \ads background. 
 Here we shall  consider  the familiar example of the Liouville theory \ci{Polyakov:1981rd,Teschner:2001rv,Nakayama:2004vk}
\be
\la{1.1}
S = \frac{1}{4\pi}\int d^{2}x\,\sqrt{g}\, \big(\partial^{a}\varphi\partial_{a}\varphi
+\mu\,e^{2\,b\,\varphi}+Q\,R\,\varphi\big),\qquad\qquad  Q=b+b^{-1} \ . 
\ee
Defined on a fixed curved 2d background  it is  a
  Weyl-covariant quantum theory with the central charge $c=1+6\,Q^{2}$.
In the special case of the  Euclidean \ads   background $ds^{2} = \frac{1}{\bz^{2}}(dt^{2}+d\bz^{2})$
 with the curvature $R= -2$ the  field $\vp$  expanded near its    constant vacuum  
 value has the classical ($b \ll 1$)  
 mass $m^2= 2$.  
 Interpreting this model from the point of   view of  the \adsf duality\foot{This approach 
is    different  in spirit from  the earlier investigations  in \ci{DHoker:1983msr}  and \ci{Zamolodchikov:2001ah,Menotti:2004uq}.}
 and assuming 
 the   standard (Dirichlet)  boundary condition at $\bz=0$   this field  should 
 have the  boundary asymptotics 
 $\varphi(t, \bz)\big|_{\bz\to 0} =\bz^2 \Phi (t) + ...$, i.e.  should 
  be    dual   to  a 1d boundary CFT operator  with the 
   1d conformal  dimension $\Delta=2$.\foot{This  is the special case of the  familiar 
    AdS$_{d+1}$ relation   $m^2= \Delta (\Delta-d)$.}
    One can then compute the corresponding boundary correlators\foot{Here
    the l.h.s.  may be viewed as a symbolic notation for the corresponding boundary CFT 
    correlator.}
    \be \la{1.2}
    \langle \Phi(t_{1})\cdots \Phi(t_{n})\rangle
    \equiv \lim_{\bz_1,...,\bz_n\to 0} \bz^{-2}_1 ...\bz^{-2}_n
    \, \langle\varphi(t_{1}, z_{1})\cdots\varphi(t_{n}, z_{n})\rangle
    \ee
  using the 
 standard Witten diagram prescription. 
 
 Given the  special  conformal invariance  
    properties  of the Liouville   theory  one may  expect these  boundary 
 correlators  to have a particularly simple   structure.
  Indeed, it was recently  noticed in  \cite{Ouyang:2019xdd}  that at  the 
 tree-level ($b\ll 1$ or $c\gg 1$)    the  2-, 3-  and 4-point  boundary correlators \rf{1.2}
   have exactly the same dependence on the 
 boundary points $t_i$ as   the correlators of the  holomorphic stress tensor $T(z)$  (generator of the Virasoro algebra with the same central charge $c$) have 
 on the complex coordinates $z_i$, i.e. 
  \be
\la{1.3}
\langle \Phi(t_{1})\cdots \Phi(t_{n})\rangle = \kappa^{n}\, \langle T(z_{1})\cdots T(z_{n})\rangle\Big|_{z_{i}\to t_{i}} \ . 
\ee
Here $ \kappa=\kappa(b) $ is a  proportionality  coefficient in the formal  identification 
$\Phi(t) \to  \kappa\, T(t)$. Equivalently,  one may view the r.h.s. of  \rf{1.3} 
as the correlator  of the chiral stress tensor of a  
CFT on a half-plane restricted to the real  boundary $z_i=t_i + i y_i \to t_i$
(assuming the usual  conformal gluing condition $T(t) = \bar T( t)$).
Explicitly, the Virasoro  algebra   or the  standard  OPE, 
  $T(z') T(z)= {c/2\ov (z'-z)^4} +
{ 2 T(z) \ov (z'-z)^2} + { \del_{z} T(z) \ov z'-z} + ...$, 
fixes the correlators  of  the stress tensor  $T(z)$  to be 
  \begin{align}
\la{1.4}
\langle & T(z_{1})\, T(z_{2})\rangle = \frac{c}{2\,z_{12}^{4}}, \qquad\qquad \qquad
\langle T(z_{1})\, T(z_{2})\, T(z_{3})\rangle =\frac{c}{z_{12}^{2}\,z_{13}^{2}\,z_{23}^{2}},\ \\
\langle & T(z_{1}) \, T(z_{2}) \, T(z_{3})\,  T(z_{4})\rangle = \frac{c^{2}}{4}\,\Big(\frac{1}{z_{12}^{4}\,z_{34}^{4}}
+\frac{1}{z_{13}^{4}\,z_{24}^{4}}+\frac{1}{z_{14}^{4}\,z_{23}^{4}}\Big)\notag \\
&\qquad\qquad \qquad \qquad \qquad \quad+c\,\Big(
\frac{1}{z_{12}^{2}\,z_{23}^{2}\,z_{34}^{2}\,z_{14}^{2}}
+\frac{1}{z_{13}^{2}\,z_{24}^{2}\,z_{14}^{2}\,z_{23}^{2}}
+\frac{1}{z_{12}^{2}\,z_{24}^{2}\,z_{34}^{2}\,z_{13}^{2}}
\Big).\la{1.5}
\end{align}
The  fact that the boundary  operator dual to the Liouville
  field $\vp$  should have the  dimension $2$   is indeed consistent with the structure 
of \rf{1.4},\rf{1.5}.\footnote{$T$   has indeed 
dimension 2 with respect to the   $SL(2,\mathbb R)$  subgroup 
of the 2d conformal  group.
Note also that $\langle T\rangle=0$ by the conformal symmetry in the bulk
so one needs to  consider only the connected correlators  or 
subtract  the expectation value 
 $\vp\to \vp-\langle\vp\rangle$  in (\ref{1.2}) 
to make the identification (\ref{1.3}) possible.}

Our aim will be to check the conjecture \rf{1.3} beyond the leading tree level 
approximation discussed in \cite{Ouyang:2019xdd}
by directly  computing  the one-loop   corrections to the  correlators $\langle \Phi(t_{1})\cdots \Phi(t_{n})\rangle$ in \rf{1.2} 
starting  with  the Liouville action \rf{1.1}  in AdS$_2$.
We shall also  discuss    how  one may try to prove 
 the relation \rf{1.3}   general  and suggest the exact
 expression for $\k(b)$. 

One  prediction  of  the identification 
 \rf{1.3}  is that the boundary operator  dual to $\vp$  should have no anomalous dimension, i.e. 
the two-point and three-point functions \rf{1.2}  should  be  given by 
 \be
\la{1.6}
\langle \Phi(t_{1})\,\Phi(t_{2})\rangle = \frac{C_{2}(b)}{t_{12}^{4}}, \qquad\qquad \qquad 
\langle \Phi(t_{1})\,\Phi(t_{2})\,\Phi(t_{3})\rangle = \frac{C_{3}(b)}{t_{12}^{2}\,t_{13}^{2}\,t_{23}^{2}} \ , 
\ee
i.e. any  logarithmic corrections  should cancel out.
The consistency with  \rf{1.3},\rf{1.4}   implies 
\be \la{1.7} 
C_2 = \ha c\, \k^2 \ , \qquad    C_3 = c\, \k^3 \ , \qquad \qquad 
 \frac{ (2\, C_{2})^{3}}{(C_{3})^{2}} = c \ , \qquad  \frac{ C_3}{2\, C_2} = \k \ .  \ee
 Since  $\k(b)$  is the only a priori  unknown function 
 (the central charge  is assumed to be given by the  Liouville theory value  
 $c= 1 + 6 ( b + b^{-1})^2= 6 b^{-2} + ...$)
    we thus get  non-trivial  relations   between  the 
  coefficients  in the perturbative  expansion of the functions 
 $C_2$ and $C_3$
 \be
\la{1.8}
C_{2}(b) = C_{2,0}+C_{2,1}\,b^{2}+C_{2,2}\,b^{4}+\cdots, \qquad
C_{3}(b) = C_{3,0}\,b+C_{3,1}\,b^{3}+C_{3,2}\,b^{5}+\cdots,
\ee
in particular, 
\be
\la{1.9}
\text{(i)}
\ \ \frac{4\,(C_{2,0})^{3}}{(C_{3,0})^{2}}=3,\qquad \qquad\qquad
\text{(ii)}
\ \ \frac{8\,(C_{2,0})^{2}}{(C_{3,0})^{3}}(3\,C_{2,1}C_{3,0}-2\,C_{2,0}C_{3,1}) = 13.
\ee
The   tree-level relation (i)  in \rf{1.9}   was  
 checked  in \cite{Ouyang:2019xdd}   and  we will  show 
 that  the  one-loop  relation (ii)  is also   satisfied. 

 It was also observed in \cite{Ouyang:2019xdd}  that the 4-point boundary correlator given  by the sum of the disconnected and connected 
  tree-level diagrams
  following from the Liouville action \rf{1.1}, i.e. symbolically
 \begin{align}
\la{1.10}
\langle\Phi\Phi\Phi\Phi\rangle &= \Big[
\begin{tikzpicture}[line width=0.4 pt, scale=0.3,baseline=-0.1cm]
\coordinate (A1) at (140:2);
\coordinate (A2) at (-140:2);
\coordinate (A3) at (40:2);
\coordinate (A4) at (-40:2);
\draw (A1) to[out=0,in=0] (A2);
\draw (A3) to[out=180,in=180] (A4);
\draw[densely dashed] (0,0) circle (2);
\draw[fill=black] (A1) circle (0.12); 
\draw[fill=black] (A2) circle (0.12); 
\draw[fill=black] (A3) circle (0.12); 
\draw[fill=black] (A4) circle (0.12); 
\end{tikzpicture}\ \ +\text{crossed}\Big] 
+b^{2}\,\Big[
\begin{tikzpicture}[line width=0.4 pt, scale=0.3,baseline=-0.1cm]
\coordinate (A1) at (140:2);
\coordinate (A2) at (-140:2);
\coordinate (A3) at (40:2);
\coordinate (A4) at (-40:2);
\coordinate (M1) at (-1,0);    \coordinate (M2) at (1,0);
\draw[densely dashed] (0,0) circle (2);
\draw (A1)--(0,0)--(A2); \draw (A3)--(0,0)--(A4); 
\draw[fill=black] (A1) circle (0.12); 
\draw[fill=black] (A2) circle (0.12); 
\draw[fill=black] (A3) circle (0.12); 
\draw[fill=black] (A4) circle (0.12); 
\end{tikzpicture}\  + \ 
\begin{tikzpicture}[line width=0.4 pt, scale=0.3,baseline=-0.1cm]
\coordinate (A1) at (140:2);
\coordinate (A2) at (-140:2);
\coordinate (A3) at (40:2);
\coordinate (A4) at (-40:2);
\coordinate (M1) at (-1,0);    \coordinate (M2) at (1,0);
\draw[densely dashed] (0,0) circle (2);
\draw (A1)--(M1)--(A2); \draw (A3)--(M2)--(A4); \draw (M1)--(M2);
\draw[fill=black] (A1) circle (0.12); 
\draw[fill=black] (A2) circle (0.12); 
\draw[fill=black] (A3) circle (0.12); 
\draw[fill=black] (A4) circle (0.12); 
\end{tikzpicture}\ + \text{crossed}
\Big]+\mc O(b^{4})
\end{align}
has a  structure  which 
is  consistent with \rf{1.3},\rf{1.5}. 
 In general, one should  expect to find
 \begin{align}
\la{1.11}
\langle \Phi(t_{1})\cdots \Phi(t_{4})\rangle
  &= \langle \Phi(t_{1})\cdots \Phi(t_{4})\rangle_{\rm disconn}  + \langle \Phi(t_{1})\cdots \Phi(t_{4})\rangle_{\rm conn}  \ , \\
\langle \Phi(t_{1})\cdots \Phi(t_{4})\rangle_{\rm disconn} &= [C_{2}(b)]^2\,\Big(
\frac{1}{t_{12}^{2}\,t_{34}^{2}}
+\frac{1}{t_{13}^{2}\,t_{24}^{2}}
+\frac{1}{t_{14}^{2}\,t_{23}^{2}}\Big) \ ,  \la{1.12} \\
\langle \Phi(t_{1})\cdots \Phi(t_{4})\rangle_{\rm conn} &= C_{4}(b)\,\Big(
\frac{1}{t_{12}^{2}\,t_{23}^{2}\,t_{34}^{2}\,t_{14}^{2}}
+\frac{1}{t_{13}^{2}\,t_{24}^{2}\,t_{14}^{2}\,t_{23}^{2}}
+\frac{1}{t_{12}^{2}\,t_{24}^{2}\,t_{34}^{2}\,t_{13}^{2}}
\Big), \la{1.13} \\
C_{4}= c\, \k^4 &= C_{4,0}\,b^{2}+C_{4,1}\,b^{4}+C_{4,2}\,b^{6}+\cdots\ .  \la{1.14}
\end{align}
Using \rf{1.7},\rf{1.9}  to eliminate $\k$ we  should  then  have  the following relations  for the coefficients  in the perturbative expansion
of $C_4$  in \rf{1.14}
\be
\la{1.15}
\text{(iii)} \ \ C_{4,0} =\frac{(C_{3,0})^{2}}{2\,C_{2,0}}, \qquad\qquad 
\text{(iv)} \ \ C_{4,1} = \frac{C_{3,0}}{2\,(C_{2,0})^{2}}\,\big(2\,C_{2,0}C_{3,1}-C_{2,1}C_{3,0}\big) \ . 
\ee
The  tree-level  relation  (iii) was checked  in \cite{Ouyang:2019xdd}  while 
 the non-trivial one-loop relation (iv) will be  confirmed   below. 
 
Our one-loop results will thus 
    provide   a   strong evidence for the validity  of the  conjecture \rf{1.3}. 
One   may wonder how it can be proved in general.  
The conformal symmetry of the Liouville theory in the bulk 
(on half-plane) viewed as an infinite  global  symmetry extending the  
 $SL(2,\RR)$       isometry of  the  AdS$_2$  
 implies strong constraints on the  boundary correlators  which are effectively equivalent to the Virasoro symmetry constraints  
 on the form of the  stress tensor  correlators \rf{1.4},\rf{1.5}. 
 This  suggests that  here we get   an 
 effective  CFT$_1$  which is  
 simply a restriction of a CFT$_2$  defined on a  half-plane to its real-line boundary
 so that in this case 
  the \adsf    becomes   effectively  "AdS$_2$/$(CFT_{2})^{1/2}$
 duality".\foot{This may be compared to  the    discussion  
of  the conjectured  duality  between a gravitational theory in AdS$_2$  and a 
chiral  half of a 2d CFT in \ci{Strominger:1998yg}. 
  While  here we assume  that  the AdS$_2$  background is 
   fixed  and the Virasoro symmetry  is not gauged,  in 
  the  gravitational  context the 2d diffeomorphisms is a gauge symmetry
  and  the Virasoro symmetry (with $c=0$ due to ghosts, etc.) 
  was  assumed to be   an  asymptotic symmetry
  (also, the Liouville field in \ci{Strominger:1998yg}  was a  gravitational mode). 
  Whether this symmetry is  spontaneously broken  to 
  $SL(2,\mathbb R)$ appears to depend on a particular model  in question
  (cf. \ci{Almheiri:2014cka,Maldacena:2016upp,Engelsoy:2016xyb}).
  }

 In the semiclassical limit of small $b$  the 
  relation \rf{1.3}  may then  be argued  for as follows.  
Since the theory \rf{1.1} is  Weyl-covariant theory  in the bulk,  
in this limit  
 we may  eliminate   the  conformal factor of the AdS$_2$  metric by  redefining  the Liouville  field   as  
 $\varphi(t,\bz) \to  \phi(t,\bz) + {b}^{-1} \log\bz$, i.e. transform  the  action in (\ref{1.1}) into the flat-space one
  (cf.  \ci{DHoker:1983msr,Zamolodchikov:2001ah}).
 This gives  
 the  Liouville CFT  for field $\phi$  defined on a   flat upper half-plane ($w=t+i\,\bz, \  \bz>0$)
 %
 %
 with the  (classical) stress  tensor $T(w) = -(\partial_w\phi)^{2}+b^{-1}\,\partial^{2}_w\phi$
 (where $\partial_{w} = \frac{1}{2}(\partial_{t}-i\,\partial_{\bz})$).
 The shifted field has then the  
 boundary asymptotics $\phi(t,\bz)\big|_{\bz\to 0} =\bz^{2}\,\Phi(t) -{b}^{-1}\log\bz+\dots$.
 Taking the boundary 
 limit in $T$  ($w\to t$) gives 
 \be
 \la{1.16} 
 T(t)  =-\tfrac{3}{2}\,b^{-1}\,\Phi(t) + \mc O(\bz^{2}).
 \ee
 This is  precisely  the operator relation   which is required  for the validity of \rf{1.3} 
 with $\kappa(b) = -\frac{2}{3}\,b+\cdots$. The same  value of  the leading term 
 in $\k(b) $ in \rf{1.3},\rf{1.7}   will   be found    by an  explicit  computation  below  (see \rf{3.22}).  

 This argument should admit an extension  to  the full quantum level, i.e. to 
 all orders in $b$,  allowing to determine   an  exact expression for $\k(b)$. 
 One important point is    the role of the boundary limit:
  the relation \rf{1.3} between  the elementary Liouville  field 
 correlators and the stress tensor correlators  may   hold  only 
 at the boundary.  
 A general   proof   should    use  the  boundary conformal field theory 
 considerations.
   In particular,  the completeness  of the OPE in the boundary CFT (cf. Appendix \ref{app:tttt}) 
 implies that it contains only the boundary stress tensor  and its  descendants; 
 since the boundary operator dual to the $m^2 =2$ Liouville field  in \ads 
 should have dimension 2  it can only be
  (proportional to)  the stress tensor itself.  Considering 
    the   OPE of the  stress tensor  in the bulk and the Liouville   field at the boundary 
    should   then 
    fix the value of the proportionality constant $\kappa(b)$ or $C_2(b)$ in the two-point 
    function in \rf{1.6}.\foot{We  are grateful to 
     Simone  Giombi and Xi Yin for  useful  suggestions  on  this problem.}
 
A possible   generalization of the above   semiclassical  argument is as follows. 
Starting with the Liouville  theory on   a  complex plane, 
  the (normal ordered) operator $V_{a}=e^{2\,a\,\phi}$ is a primary 
with the conformal dimension $\Delta=a\,(Q-a)$, i.e.   we have 
\be
T(w)\,V_{a}(w', \bar w') = \frac{a(Q-a)}{(w-w')^{2}}\,V_{a}(w',\bar w') + \frac{1}{w-w'}\,\del_{w'}V_{a}(w', \bar w') +\cdots
 \ . 
\ee
Expanding this  formally  in small $a$  gives 
$
T(w)\,\phi(w',\bar w')\sim  \frac{Q}{2\,(w-w')^{2}}+\frac{\del_{w'}\phi(w',\bar w')}{w-w'}.
$
On the upper half-plane ($w=t + i \bz, \ \bz>0$) 
  the boundary conformal Ward identity takes into account 
also the poles at $w=\overline{w}'$. Including them in the OPE of $T$ and $\phi$ 
and 
assuming  the  exact form of the boundary  asymptotics to be\foot{This  generalizes  the classical asymptotics 
where $Q\to b^{-1}$ in a way which  is consistent with the value of the one-loop tadpole   (see \ci{Zamolodchikov:2001ah}  and   \rf{21.7}).}
\be \la{1188}
\phi(t,\bz)\big|_{\bz\to 0} =\bz^{2}\,\Phi(t) -  Q \log\bz+\dots\ , \ee
\iffa
using again
 the boundary asymptotics $\phi(t,\bz)\big|_{\bz\to 0} =\bz^{2}\,\Phi(t) -{b}^{-1}\log\bz+\dots$
we find for  the following most singular term (the coefficient of the identity operator)
\be\la{1.18} 
T(t)\Phi(t') = 
\frac{1}{\bz'^{2}}\frac{Q- b^{-1}}{(t-t')^{2}}-\frac{3Q- b^{-1}}{(t-t')^{4}}+\cdots.
\ee
where dots stand for higher dimensional operators and higher orders in $\bz'\to 0$.
Like in the semiclassical expression 
\rf{1.16} here the  first term  vanishes  at the leading order in small $b$  expansion when   $Q\to b^{-1}$  so \rf{1.18}. As this term  should   vanish for all $b$ 
we expect  that a refined version of  the relation \rf{1.18}  should   have 
$b^{-1}$ terms  replaced by $Q= b^{-1} + b$.  
\fi
 one finds that the leading $1\ov (t-t')^2$ term in $T(t)\phi(t,0)$  cancels and
we get\footnote{In more details, Eq. (\ref{120}) is the coefficient of $\bz^{2}$ term in the $\bz\to 0$ expansion of 
$\frac{Q}{2\,(t-t'-i\,\bz)^{2}}+\frac{i}{2}\,Q\frac{1}{\bz}\frac{1}{t-t'-i\,\bz}+\text{c.c}$ where  the second term comes from the $-Q\log\bz$ term in (\ref{1188})
and the complex conjugate 
contribution comes from  poles at $w=\overline{w}'$.
}
\be\la{120}
T(t)\,\Phi(t') =  - \, 
\frac{2Q}{(t-t')^{4}}+\cdots  \ . 
\ee
Taking the expectation value of \rf{120} 
  and comparing  the result  with \rf{1.4},\rf{1.6},\rf{1.7} we get 
 the following  prediction  for  
the  exact value of the  coefficient $\k(b)$   between $\Phi$ and $T$:
%
\begin{align}
\la{911}
&\Phi = \kappa\,   T \ , \qquad \qquad \qquad \qquad  \kappa(b)  = -\frac{4Q}{c} \ , \\
\la{119}
&\kappa(b)   = -\frac{4Q}{1 + 6 Q^2 }
 = - \frac{ 4\,b  (1 + b^2)}{(3+2\,b^{2})(2+3\,b^{2})} = 
-\frac{2}{3}\,b+\frac{7}{9}\,b^{3}
-\frac{55}{54}\,b^{5}+\cdots\ . 
\end{align}
The first  two terms in this expansion  match precisely 
  with the above   semiclassical value and the one-loop 
value  that we shall find below in  (\ref{3.24}). 
Given \rf{911}  we get from \rf{1.7} 
\be \la{1.77} 
C_2 = {8 Q^2 \ov c} = {4 \ov 3} - {2 \ov 9}  b^2  +...    \ , 
\qquad  \qquad   
 C_3 = -{64 Q^3 \ov c^2} = -{16 \ov 9}b  + {64 \ov 27}  b^3  +... \ , 
  \ee
which are in agreement with  \rf{3.259}.

\iffa
Comparing the second term with \rf{1.4},\rf{1.6},\rf{1.7} we find ($c\sim\frac{6}{b^{2}}+\cdots$)
\be
\la{k2}
\k(b) = \frac{2}{c}\frac{1-3\,b\,Q}{b}=-\frac{2}{3}\,b+\mc O(b^{3}),
\ee 
in agreement with the previous calculation in (\ref{1.16}). 
One can check that the cubic corrections differ; one finds $\kappa=-\frac{2}{3}b+\frac{2}{9}b^{3}+\cdots$
from (\ref{1.16}) and $\kappa=-\frac{2}{3}b+\frac{4}{9}b^{3}$ from (\ref{k2}). This is not surprising
because \red{WHY? What can we quote as the reason for the above arguments not being exact ?}.
We shall see later that the one-loop correction to $\kappa(b)$ turns out to be $+\frac{7}{9}\,b^{3}$,
see below  (\ref{3.24}),
suggesting that an exact expression could be the simple ratio 
$\kappa(b) = -4\,Q/c = 
-\tfrac{2}{3}\,b+\tfrac{7}{9}\,b^{3}
-\tfrac{55}{54}\,b^{5}+\cdots$
\fi

%
%
%


 A natural generalization 
  of the  above  duality \rf{1.3}   for the Liouville theory 
 is to  all   conformal Toda theories  
 for  finite Lie algebras \cite{Gervais:1983am,Mansfield:1982sq,Braaten:1983pz}
 defined on  the   AdS$_{2}$   background.
 For instance, in the $A_{n}$ case,  expanding near the minimum of the Toda potential 
 in AdS$_{2}$ one finds  $n$ scalar fields $\varphi_{\de}$ with masses $m^{2}=\de(\de-1)$   corresponding to 
  $\de=2, \dots, n+1$  \cite{Ouyang:2019xdd}. 
  The duality relation extending 
  (\ref{1.3}) is then 
\be
\la{1.17}
\langle \Phi_{\de_{1}}(t_{1})\cdots \Phi_{\de_{n}}(t_{n})\rangle =\big( \prod_{i=1}^{n}\kappa_{\de_{i}}\big) \ 
\langle Q_{\de_{1}}(z_{1})\cdots Q_{\de_{n}}(z_{n})\rangle\Big|_{z_{i}\to t_{i}} \ ,
\ee
where  $Q_\de= \{Q_{2}\equiv T, Q_{3}, \dots, Q_{n+1}\}$ are the generators of the chiral $\mc W_{n+1}$ algebra  generalizing 
  the Virasoro symmetry
 and  having  the same central charge as the Toda theory in the bulk. 
The relation (\ref{1.17}) was  noticed  at the tree level in \cite{Ouyang:2019xdd} 
by computing few sample 4-point functions
in the Toda theories  associated to  some  rank-2 algebras
 with two scalar fields (one dual to the
stress  tensor $T$ and the other dual to a higher spin current $Q_{s}$).
 It is natural to expect (\ref{1.17})
to hold also at the  quantum level  as 
should be possible to check   by the methods  used  in the present paper. 
Another  interesting   generalization is to the 
 super-Liouville theory on AdS$_{2}$, cf. \cite{Ahn:2002ev}. 
 In this case one will  need to compute loop corrections
for both  the bosonic  and fermionic fields.

\


The structure of the rest of  this paper is as follows. In  section \ref{sec2.1} we 
will discuss
two alternative formulations  of the Liouville
theory on AdS$_{2}$.  One  starts with  the flat space action expanded 
around the classical solution  corresponding  effectively to the  AdS$_{2}$  
 background 
 while the other 
starts directly with  the Liouville theory   defined on  fixed   AdS$_{2}$. 
The two formulations differ
  in the  choice of the regularization of the 
short distance propagator and also in the coefficient 
in front of the potential. 
They turn out to  give the equivalent results   for 
 the physically relevant 
 connected one-loop correlators.
 
In  section   \ref{sec2.2} 
we will compute  the   two-point function  of the Liouville field 
at the one-loop order
demonstrating the consistency  of its boundary limit with \rf{1.6}.
  The one-loop  corrections to the  boundary 
three-point   and four-point functions will   be  computed in 
sections ~\ref{sec:threepoints} and \ref{sec:fourpoints} respectively
finding again the agreement with  \rf{1.6} and \rf{1.13}.
As a result, we  will  provide  a strong 
support for the  validity of the  duality  (\ref{1.3}) at the  loop level. 
Appendix \ref{app:disc} will present some  details of the 
  diagram computations on the Poincar\'e disc.
  Comments on tadpole   diagrams (that lead to different predictions 
  for the 1-point functions in the two formulations)  
  will be made in Appendix \ref{app:tadpole}.
  In Appendix  \ref{app:tttt} we  will review the structure of the 1d conformal  block 
  expansion of the four-point function in \rf{1.5}.  

\section{
Liouville theory on Euclidean   AdS$_{2}$  background}
\la{sec:tree}

The  correlators  in the Liouville  theory \rf{1.1} in AdS$_{2}$  background
may be  computed using two alternative   approaches.
In the first (the "\ZZ formulation" \ci{Zamolodchikov:2001ah}) one starts with the Liouville action on a flat upper half plane (or flat disc) 
and expands it near a non-trivial  non-constant 
 solution \cite{DHoker:1983zwg,DHoker:1983msr}  preserving the $SL(2,\RR)$  symmetry.
In the second  (the "AdS formulation")  one  starts directly  with the  Liouville   action  \rf{1.1}  in  the  \ads   background  and expands  near the  constant  minimum of the
curved-space  potential.\foot{This is  similar 
 to the  so-called  "geometrical" approach \cite{Takhtajan:1994vt}
  discussed in detail in   \cite{Menotti:2004uq}.}
 The two  approaches 
 are  classically equivalent   by a field redefinition  but  imply the  use of 
 different regularizations at the quantum level. 
 
\iffa \footnote{
For general (recent) reviews on Liouville theory see \cite{Teschner:2001rv,Nakayama:2004vk}.
Old papers mentioning the  relation of Liouville theory near
special solution to Liouville in AdS$_{2}$ are  \cite{DHoker:1983zwg,DHoker:1983msr}. 
A detailed treatment appeared later in \cite{Zamolodchikov:2001ah}. For standard renormalization group analysis 
of the UV divergences and conformal fixed points see  \cite{Grisaru:1990gf}.
}
\be
\la{2.1}
S = \frac{1}{4\pi}\int d^{2}x\,\sqrt{g}\,[\partial^{a}\varphi\partial_{a}\varphi
+\mu\,e^{2\,b\,\varphi}+Q\,R\,\varphi],\qquad Q=b+b^{-1}.
\ee
It is Weyl-invariant at the quantum level with central charge $c=1+6\,Q^{2}$. \footnote{
The relation between $b$ and $Q$ can be understood, e.g., as the condition of the vanishing
of the Weyl-anomaly coefficient of the $\beta$-function for the tachyon coupling (see, e.g. \cite{Tseytlin:1990mz}):
$\overline{\beta}^{T}=\alpha'\,(-\frac{1}{2}\,\nabla^{2}T+G^{mn}\partial_{n}\phi
\partial_{m}T+\frac{1}{2}m^{2}T)$ where in (\ref{2.1})
$T=\mu\,e^{2\,b\,\varphi}$, $G_{mn}=G_{\varphi\varphi}=1$, $\phi=\alpha'\,Q\,\varphi$, 
$m^{2}=-\frac{4}{\alpha'}$ and $\alpha'$ is set to 1. 
}
\fi 


\subsection{Two  alternative formulations  \la{sec2.1}}   

The flat-space action\foot{We shall use the notation 
$z=x_1 + i x_2$, \  $d^2z= dx_1 dx_2$.
As usual,  here  one also  assumes  that the  stress tensor of  the theory \rf{22} 
 is $T=-( \del \varphi)^2 + Q \del^2 \varphi$ 
 in order to make the interaction term exactly
marginal.}
\be\la{22}
S = \frac{1}{4\pi}\int d^{2}z\,\big(\partial^{a}\varphi\partial_{a}\varphi
+\mu\,e^{2\,b\,\varphi}\big)
\ee
\iffa
The action (\ref{2.1}) reduces in flat space to the $Q$-independent form \footnote
{The parameter $Q$ is implicitly present since in flat space we require conformal invariance with 
respect to the stress tensor with extra term $Q\partial^{2}\varphi$ and, accordingly, the boundary conditions at infinity 
$\varphi(x)\sim -Q\,\log|x|$ for $|x|\to \infty$.}
\fi
admits a particular solution  
\be
\la{2.3}
\varphi^{(0)} = -\tfrac{1}{2\,b}\log[\tfrac{\mu}{4}\,b^{2}\,(1-z\,\overline z)^{2}]\ .
\ee
Expanding near this "\ads vacuum",  
 $\varphi=\varphi^{(0)}+\chi$, we find the following action for the fluctuation $\chi$
\be
\la{2.4}
S^{\sf ZZ} =  \frac{1}{2\pi}\int d^{2}z\,\Big[\frac{1}{2}\,(\partial_{a}\chi)^{2}
+\frac{2\,(e^{2\,b\,\chi}-2\,b\,\chi-1)}{b^{2}\,(1-z\,\overline{z})^{2}}\Big]\ . 
\ee
This  action is  the same  as  the classical action \rf{1.1} (with $Q\to b^{-1}$) 
in \ads background in the Poincar\'e disc (or "pseudosphere")  coordinates  where 
 $ds^2 = {4 dz d \bar z \ov (1- z\bar z)^2}$
that has  $SU(1,1) \simeq SL(2,\RR)$   symmetry ($z\to \frac{a\,z+b}{\overline{b}\,z+\overline{a}},\ \  |a|^{2}-|b|^{2}=1$).
An alternative choice  is the half-plane parametrization
 with coordinates
$t\in\mathbb R$, $\bz\ge 0$
and the metric $ds^{2}=\frac{1}{\bz^{2}}(dt^{2}+d\bz^{2})$.
The two  are related by  the standard  conformal map 
\be
\la{2.7}
w \equiv  t+i\,\bz = i\,\frac{1+z}{1-z}\ , \ \ \ \ \ 
z=  \frac{w-i}{w+i} \ , \ \ \ \ \ \ \  
ds^2= -  {4d w d \bar w \over ( w- \bar w)^2} = {4 d z d \bar z \over ( 1- z \bar z)^2}\ . 
\ee
The  analog of the action (\ref{2.4})  on the Poincar\'e plane    reads
\be
\la{2.8}
S^{\sf ZZ} = 
\frac{1}{2\pi}\int dt\,d\bz \,
\Big[{1\ov 2}   (\partial_{a}\chi)^{2}
+\frac{e^{2\,b\,\chi}-2\,b\,\chi-1}{2 b^{2}\, \bz^2}\Big].
\ee
As follows from \rf{2.4} or \rf{2.8} the  field $\chi$ 
has  the mass  $m^2 =\de(\de-1)=2$. 
Assuming the  standard Dirichlet  boundary condition 
(corresponding to the  $\de=2$ choice) the canonically normalized 
 \ads bulk-to-bulk  propagator is given by \be
\la{2.10}
G_{\de=2} = \,\frac{\mc C_{2}}{16\,u^{2}}\,_{2}F_{1}(2, 2, 4; -\tfrac{1}{u}) = 
-\frac{1}{4\,\pi}\,\Big[(1+2\,u)\,\log\frac{u}{u+1} + 2 \Big]\ , 
\ee
where $\mc C_{2} = \tfrac{\Gamma(2)}{2\,\sqrt\pi\,\Gamma(2+\frac{1}{2})}
=  {2 \ov 3 \pi}$
and 
$u$ is the invariant  chordal distance. Equivalently (taking into account the 
factor $\frac{1}{2\pi}$ in front of the action in \rf{2.4},\rf{2.8}) 
the free  two-point function of the field $\chi$ is given by  
\begin{align}
\la{2.5}
&g(z,z') \equiv \langle \chi(z)\,\chi(z')\rangle_0    =  2\pi G_{\de=2} 
=  -\frac{1}{2}\Big(  \frac{1+\eta}{1-\eta}\log\eta + 2  \Big)\ , \\
&
\la{2.9}
\eta=  \left|\frac{z-z'}{1-z\,\overline{z}'}\right|^{2}
 =  \left|\frac{w-w'}{w-\overline{w}'}\right|^{2}= \frac{u}{u+1} \ , \ \  \ \ \qquad
u = \frac{(t-t')^{2}+(\bz-\bz')^{2}}{4\,\bz\,\bz'}.
\end{align}
 Taking  one point to the boundary  we get the corresponding bulk-to-boundary
  propagator $g_\bb$     written in either 
    disc or plane parametrization\foot{The factor $4\ov 3$ is  the same as 
  $2\,\pi\,\mc C_{2}$, where $\mc C_{2}$ appeared in \rf{2.10}.}
\be
\la{2.27}
\bz^{-2}\,g(z,z')\big|_{|z|\to 1}\equiv  g_\bb(\theta,z') = g_\bb(t,w')
= \frac{4}{3}\,\frac{\sin^{4}(\frac{\theta}{2})\,
(1-|z'|^{2})^{2}}{|e^{i\,\theta}-z'|^{4}}=\frac{4}{3}   \Big[\frac{\bz'}{\bz'^{2}+
(t-t')^{2}}\Big]^{2}\ , 
\ee
where the relation between the  Poincar\'e  plane    and  
disc   boundary coordinates
 $ t$ and $\theta$ is 
\be
\la{2.26}
 t(\theta) = i\,\frac{1+e^{i\theta}}{1-e^{i\,\theta}} = -\cot\frac{\theta}{2}\ . 
 \ee
 \iffa 
 \qquad \qquad
\begin{tikzpicture}[line width=1 pt, scale=0.3, rotate=0,baseline=0]
\draw[thin] (0,0) circle (2);
\draw[-latex,thin] (150:2.2) arc (150:120:2.2);
\draw[-latex,thin] (240:2.2) arc (240:210:2.2);
\draw[thin,-latex] (-2.5,0)--(2.5,0); \draw[thin,-latex] (0,-2.5)--(0,2.5);
\draw[fill=white] (-2,0) circle (0.12);  \node[left=0.2cm] at (-2,0) {$t=0$};
\draw[fill=white] (10:2) circle (0.12); \node[right] at (20:2) {$t\to +\infty$};
\draw[fill=white] (-10:2) circle (0.12);\node[right] at (-20:2) {$t\to -\infty$};
\end{tikzpicture}
\ee 
 \fi 
 The starting point of the  {\sf AdS} formulation is the action \rf{1.1} 
 in  unit-radius  \ads   background ($R=-2$). 
 Shifting the field  by a  constant  as  
 $\vp = \vp^{(0)} + \chi$   where $\mu e^{ 2 b \vp^{(0)} } = {Q\ov b}$  we get the following analog of \rf{2.4}
 \be
\la{2.12}
S^{\sf AdS}=  \frac{1}{2\pi}\int d^{2}z\,\Big[\frac{1}{2}\,(\partial_{a}\chi)^{2}
+\frac{2\,Q\,(e^{2\,b\,\chi}-2\,b\,\chi-1)}{b\,(1-z\,\overline{z})^{2}}\Big]\ . 
\ee
 The two coincide in the semiclassical limit of  small $b$ when $Q= b+ b^{-1} \to b^{-1}$
 but differ for finite $b$.

 The two formulations  will differ also  by the required  choice of the UV regularization
 of the propagator \rf{2.5} at the coinciding points
   but  the two differences will happen to 
 compensate  each other   leading  to the equivalent results  for  the  relevant 
 physical correlators (but  not for the tadpole values). 
 The \ZZ  formulation  that starts  with the flat-space action requires the 
  use of the simple  flat-space  
 regularization $z- z' \to z-z' + \eps$. Omitting the singular $\log \eps$ term 
  this is equivalent to the  subtraction of the  $-[\log ( z-z')]_{z\to z'}$  term   from \rf{2.5} 
 \cite{Zamolodchikov:2001ah}. At the same time, the  regularization in the AdS  formulation should   be the \ads covariant one with $\eta  \to \eta  + \eps$  or $u \to u + \eps$ in \rf{2.5},\rf{2.9}.
 In general, the   value of the  propagator at the coinciding points  may be 
 parametrized as 
 \begin{align}
 & g(z,z) = \C_{1}\, \log(1-z\,\overline{z})-\C_{2}, \la{2.13}\\
 &
\la{2.14}
\C^{\sf ZZ}_{1}=\C^{\sf ZZ}_{2}=1,\qquad  \ \ \ \ g^{\sf ZZ}(z,z) = \lim_{z'\to z}\big[g(z,z')+\log|z-z'|)\big] = \log(1-z\,\overline{z})-1\ , \\
\la{2.15}
&\C^{\sf AdS}_{1}=\C^{\sf AdS}_{2}=0,\qquad \ \ \  g^{\sf AdS}(z,z) = 0\ . 
\end{align}
Here  in \rf{2.15}   we assumed
 that the  logarithmic UV divergences  are  cancelled by the mass renormalization
   with the  "minimal" subtraction (more generally, one may  keep the constant
    $\C^{\sf AdS}_2$  arbitrary).

\iffa 
\footnote{For completeness, we remind a third formulation, the so-called 
'geometrical' approach \cite{Takhtajan:1994vt,Takhtajan:1995fd}. This approach 
is based on the introduction of localized sources in the action to compute correlators of usual  
vertex operators. At the perturbative level,
after separating out the background, the non-homogeneous source induced field, and the quantum fluctuations, 
this approach turns out to be similar to the {\sf AdS} formulation \cite{Menotti:2004uq}. 
The interaction coupling is written in terms
of  $Q= 1/b_g$, i.e. the naive coupling, while $\C_2\neq 0 $ is chosen in some ad hoc way. 
Not surprisingly, when compared with {\sf ZZ}, 
the natural identification turns out to be $\frac{1}{b_{g}} = \frac{1}{b}+b$ \cite{Takhtajan:1995fd}.
}
\fi

Starting with \rf{2.4} or \rf{2.12} one may compute the corresponding quantum corrections to 
bulk correlators in perturbation theory in $b$. The two actions differ in  the coefficient 
in front of the potential term 
\be
\la{2.16}
V^{\sf AdS} = \frac{2\,Q}{b}\,(e^{2\,b\,\chi}-2\,b\,\chi-1) = (1+b^{2})\,V^{\sf ZZ} \ , 
\ee
implying that in  the AdS  formulation one has  extra  higher order  terms in the
$\chi^2, \chi^3, ...$  vertices.  In addition, there is a  difference in the regularization
choices  in \rf{2.14},\rf{2.15}.
While the values of the one-point  functions    $\langle \chi^n(z)\rangle$
computed in the  {\sf ZZ}   and AdS  formulations
will  differ, the connected  correlators 
at separated points are  expected to be same; we shall find evidence for that below.

\subsection{One-point and two-point correlation   functions   \la{sec2.2}}

Let us  consider  the   leading order  corrections  to  the tadpole or 
the one-point function $\langle\chi(z)\rangle$. Computing it for generic choice of the regularization \rf{2.13} one finds 
(see  (\ref{A.6}),
\be
\langle \chi(z)\rangle = 
\begin{tikzpicture}[line width=1 pt, scale=0.6,baseline=-0.1cm]
\draw (-1.8,0)--(-0.8,0); \draw (0,0) circle(0.8);
\draw[fill=black] (-1.8,0) circle (0.1); 
\node[left] at (-1.8,0) {$z$};
\end{tikzpicture}  +  \mc O(b^{3})
 = b\,\Big[\tfrac{1}{2} \C_{1}+ \C_{2} -\C_{1}\,\log(1-z\,\overline z)\Big]+ \mc O(b^{3})
\ . \la{21.7}\ee
The leading-order tadpole thus vanishes in the AdS formulation
and is constant in general, as  required by the AdS symmetry (under which $\chi$ transforms as a scalar). 
Explicitly,  one finds 
\be
\la{2.18}
\langle \chi(z)\rangle^{\sf ZZ} = b\,\Big[\frac{3}{2}-\log(1-z\,\overline z)\Big]+
b^{3}\,\Big(\frac{\pi^{2}}{6}-\frac{13}{12}\Big)+\mc O(b^{5}),\qquad\qquad 
\langle \chi(z)\rangle^{\sf AdS} = \mc O(b^{3}).
\ee
Note  that combining the one-loop tadpole with  the classical 
solution \rf{2.3} the $z$-dependent part of the tadpole in the {\sf ZZ}  case   gets 
coefficient $Q= b + b^{-1}$. The $b^3$  term in $\langle \chi(z)\rangle^{\sf ZZ}$ 
is in agreement with the results of \cite{Zamolodchikov:2001ah} and 
 \cite{Menotti:2003km}  (see also Appendix \ref{app:tadpole}).\foot{The tadpole (\ref{2.18}) is the simplest cumulant that can be extracted from the expectation value of the  exponential 
vertex operator   $\langle e^{\alpha\,\chi(z)}\rangle$  determined   in \cite{Zamolodchikov:2001ah}. 
For a perturbative discussion of the $\mc O(b^{3})$ corrections
to higher cumulants and a comparison with the 
bootstrap predictions see  \cite{Menotti:2003km}.}  

Next, let us consider the one-loop  order $b^2$ 
correction to the connected part of the two-point function 
$\langle \chi(z)\chi(z')\rangle$.
In the  AdS   formulation one needs to take into account   an extra contribution coming from the $b^2\chi^2$ term  in the potential in \rf{2.16}. The final result in both formulations   can be written as 
\begin{align}
&\langle  \chi(z)\chi(z')\rangle_{\rm conn}=
\la{2.20}
\begin{tikzpicture}[line width=1 pt, scale=0.6, baseline=-0.1cm]
\draw (-1.2,0)--(1.2,0);
\draw[fill=black] (-1.2,0) circle (0.1); 
\draw[fill=black] (1.2,0) circle (0.1); 
\end{tikzpicture}
+
\begin{tikzpicture}[line width=1 pt, scale=0.6, baseline=-0.1cm]
\draw (-1.8,0)--(-0.8,0); \draw (0,0) circle(0.8);
\draw (1.8,0)--(0.8,0); 
\draw[fill=black] (-1.8,0) circle (0.1); 
\draw[fill=black] (1.8,0) circle (0.1); 
\end{tikzpicture}
+ \C_1'  
\begin{tikzpicture}[line width=1 pt, scale=0.6]
\draw (-1.5,0)--(1.5,0); 
\begin{scope}[shift={(0,0)},scale=0.6]
	   \draw[fill=white,thin] (0,0) circle (0.3);
            \draw[thin] (45:0.3)--(225:0.3);
            \draw[thin] (-45:0.3)--(135:0.3);
	\end{scope}
\draw[fill=black] (-1.5,0) circle (0.1); 
\draw[fill=black] (1.5,0) circle (0.1); 
\end{tikzpicture} +\mc O(b^{4})\\
&\qquad \qquad \qquad \ \ =g(z,z') +   \Sigma(z,z')  \no 
 \\
&+\frac{2}{3} b^2 \, \C_1' 
\,\Big[
1-\frac{\eta\,\log\eta}{1-\eta}-\log(1-\eta)\,\Big(1+\frac{(1+\eta)\,\log\eta}{2\,(1-\eta)}\Big)
-\frac{1+\eta}{1-\eta}\,\text{Li}_{2}(1-\eta)
\Big] +\mc O(b^{4}), \la{217}
\end{align}
where (see \rf{A.11},\rf{a15},\rf{A.17}) 
\be
\la{2.21}
\Sigma(z,z') 
= b^{2}\,\Big[
\frac{3}{2}+\frac{\eta^{2}\log^{2}\eta}{2\,(1-\eta)^{2}}-\frac{1+\eta}{1-\eta}\,\text{Li}_{2}(1-\eta)\Big]\ , 
\ee
and  the coefficient $\C_1'= \C_1 - 1 $ 
 in the \ZZ case and $\C_1'= \C_1  $   in  the  AdS  case.
As follows  from \rf{2.14},\rf{2.15}   in both cases 
\be   \C'_1=0 \ , \la{111}\ee
 and  so  we get the equivalent result for \rf{2.20} 
 given by the sum of  the free  propagator in \rf{2.5} and the "self-energy" correction in 
 \rf{2.21}.

Let us consider the Poincar\'e plane parametrization   and introduce the 
subtracted and rescaled   field 
\be\la{221}
\Phi(t, \bz) = \bz^{-2}\,\big[\chi(t, \bz)-\langle \chi(t, \bz)\rangle\big]\ . 
\ee
  Using (\ref{2.5}),(\ref{2.9}),(\ref{2.21})  and taking the limit $\bz_1,\bz_2\to0$ 
   we get
  for the two-point boundary correlator in \rf{1.2} (cf. \rf{2.27})
\be
\la{2.25}
\langle\Phi(t) \, \Phi( t')\rangle \equiv  \lim_{\bz, \bz'\to0} \langle  \chi(t,\bz)\, \chi(t',\bz')\rangle_{\rm conn} = 
\frac{C_2(b)}{(t-t')^{4}}\ ,\qquad \qquad 
C_2= \frac{4}{3}\Big( 1 -\frac{1}{6}\,b^{2}\Big)+\mc O(b^{4})\ .
\ee
Thus there are  no $b^{2}\log(t-t')$ corrections, i.e.  the boundary  operator  dual to $\chi$ has no anomalous   dimension  confirming   the expected  behaviour 
\rf{1.6}.
We conclude that the  coefficients  in the perturbative expansion of $C_2$ introduced in \rf{1.8} are thus 
\be
\la{2.28}
C_{2,0} = \frac{4}{3}, \qquad\qquad \qquad  C_{2,1} = -\frac{2}{9}.
\ee
Let us note that it is only the boundary limit of the 2-point function \rf{2.20}  that takes the simple 
rational form in \rf{2.25} matching the stress tensor two-point function 
 \rf{1.4}  according to \rf{1.3}. The same  will apply to higher-point  correlation functions.

\iffa 
Then  the relations in \rf{1.9}  lead to the following predictions for the coefficients in the three-point  correlator in \rf{1.8}
\be
\la{2.29}
|C_{3,0}| = \frac{16}{9}\ , \ \qquad \qquad \qquad 
|C_{3,1} |= \frac{64}{27} \   
\ee
that will be confirmed below.
\fi

\section{Three-point boundary correlation function}
\la{sec:threepoints}

Next, let us  compute the  tree-level and one-loop  contributions  to the 
 connected boundary three-point function. 
 We shall use  mainly  the {\sf ZZ}
formulation, but will comment on its  equivalence with the {\sf AdS} one. 
The relevant terms in the expansion  of the potential term  in \rf{2.4} will be 
\be\la{31}
 \int  \mathsf{d}^{2}z\   \frac{e^{2b \chi}-2b\chi-1}{b^2}
 = \int  \mathsf{d}^{2}z\ \Big[  
  2{\chi^{2}}+ 8\,b\,\tfrac{\chi^{3}}{3!}+16\,b^{2}\,\tfrac{\chi^{4}}{4!}
+32\,b^{3}\,\tfrac{\chi^{5}}{5!}
+\cdots\Big]  \ , \qquad  \mathsf{d}^{2}z  \equiv  \frac{d^2 z }{ \pi (1-|z|^{2})^{2}} \ . \ee
We will  need  the tree boundary-to-bulk propagator in the disc parametrization
given in \rf{2.27}  and also will use   the  following   notation  
for the  conformal prefactor in the  three-point function \rf{1.6} (cf. (\ref{2.26}))
\be
\la{3.3}
\mc K_{3}(\bm\theta) \equiv  \mc K(\theta_{1}, \theta_{2}, \theta_{3}) = \prod_{i<j}^{3}
\big| {t}(\theta_{i})-{t}(\theta_{j})\big|^{-2}.
\ee
We shall divide  the  relevant  Witten diagrams in \ads 
by the overall conformal factor   $\mc K_{3}$, i.e.  the resulting expressions  
will represent  the  contributions to the  coefficients in $C_3(b)$ in \rf{1.6}. 

The tree level contribution to the three-point function is given by the  diagram
\begin{align}
\la{3.4}
C_{3,0}\,b &= \begin{tikzpicture}[line width=1 pt, scale=0.4, rotate=0,baseline=-0.1cm]
\coordinate (A1) at (90:2);  \coordinate (A2) at (210:2);  \coordinate (A3) at (-30:2);
\coordinate (B1) at (90:1);  \coordinate (B2) at (210:1);  \coordinate (B3) at (-30:1);
\draw[densely dashed] (0,0) circle (2);
\draw (A1)--(0,0); \draw (A2)--(0,0); \draw (A3)--(0,0);
\draw[fill=black] (A1) circle (0.1); \draw[fill=black] (A2) circle (0.1); \draw[fill=black] (A3) circle (0.1); 
\end{tikzpicture} \ \ =  \frac{(-8b)}{{\mc K}_{3}(\bm\theta)}\int \mathsf{d}^{2}z \,
\prod_{i=1}^{3} g_{\rm b}(\theta_{i}, z) 
= 
(-8\,b)\,(\tfrac{4}{3})^{3}(\tfrac{1}{4\pi})\times\tfrac{3\pi}{8} = -\tfrac{16}{9}\,b\ .
\end{align}
The same  result in the Poincar\'e plane  coordinates follows  from 
(see, e.g., \cite{Freedman:1998tz})\foot{See  \rf{2.27}   and  note  that $\frac{dt\,d\bz}{\bz^{2}}=\frac{4d^2 z }{(1-|z|^{2})^{2}}$.}
\be
\la{3.5}
\int\frac{dt\,d\bz}{\bz^{2}} \prod_{i=1}^{3}\Big[\frac{\bz}{\bz^{2}+
(t-t_{i})^{2}}\Big]^{2} = \frac{3\,\pi}{8}\,\frac{1}{|t_{12}|^{2}\
|t_{13}|^{2}\ |t_{23}|^{2}}.
\ee
Thus  the   value of the   leading coefficient  in $C_3$ in \rf{1.6},\rf{1.8}
is 
\be
\la{2.29}
C_{3,0} = -  \frac{16}{9}\ , \ee
which together with $C_{2,0}$ in \rf{2.28}  is in agreement  with the first relation in \rf{1.9}.

The value  (\ref{2.29})  combined  with \rf{2.28}    and  the second relation in 
\rf{1.9}     gives   the following 
prediction for  the one-loop coefficient  $C_{3,1}$  in   $C_3(b) $  in \rf{1.8}
\be
\la{3.6}
C_{3,1} = \frac{64}{27}.
\ee 
This   coefficient   should  be   given by the sum of the contributions of the four different types of diagrams
\be
C_{3,1} = \sum_{i=1}^{4}C_{3,1}^{(i)},\la{3.66}
\ee
that   we shall   consider  below. 

\paragraph{Diagrams  with dressed propagators:} 
There are  three tree-like  diagrams with the one-loop self-energy correction to one of the propagators
\be\la{3.7}
C_{3,1}^{(1)}\,b^{3} = \ \ 
\begin{tikzpicture}[line width=1 pt, scale=0.4, rotate=0,baseline=-0.1cm]
\coordinate (A1) at (90:2);  \coordinate (A2) at (210:2);  \coordinate (A3) at (-30:2);
\coordinate (B1) at (90:1);  \coordinate (B2) at (210:1);  \coordinate (B3) at (-30:1);
\draw[densely dashed] (0,0) circle (2);
\draw (A1)--(0,0); \draw (A2)--(0,0); \draw (A3)--(0,0);
\draw[fill=lightgray] (0,1) circle (0.5);
\draw[fill=black] (A1) circle (0.1); \draw[fill=black] (A2) circle (0.1); \draw[fill=black] (A3) circle (0.1); 
\end{tikzpicture}\ \ +
\begin{tikzpicture}[line width=1 pt, scale=0.4, rotate=0,baseline=-0.1cm]
\coordinate (A1) at (90:2);  \coordinate (A2) at (210:2);  \coordinate (A3) at (-30:2);
\coordinate (B1) at (90:1);  \coordinate (B2) at (210:1);  \coordinate (B3) at (-30:1);
\draw[densely dashed] (0,0) circle (2);
\draw (A1)--(0,0); \draw (A2)--(0,0); \draw (A3)--(0,0);
\draw[fill=lightgray] (-30:1) circle (0.5);
\draw[fill=black] (A1) circle (0.1); \draw[fill=black] (A2) circle (0.1); \draw[fill=black] (A3) circle (0.1); 
\end{tikzpicture}\ \ +
\begin{tikzpicture}[line width=1 pt, scale=0.4, rotate=0,baseline=-0.1cm]
\coordinate (A1) at (90:2);  \coordinate (A2) at (210:2);  \coordinate (A3) at (-30:2);
\coordinate (B1) at (90:1);  \coordinate (B2) at (210:1);  \coordinate (B3) at (-30:1);
\draw[densely dashed] (0,0) circle (2);
\draw (A1)--(0,0); \draw (A2)--(0,0); \draw (A3)--(0,0);
\draw[fill=lightgray] (210:1) circle (0.5);
\draw[fill=black] (A1) circle (0.1); \draw[fill=black] (A2) circle (0.1); \draw[fill=black] (A3) circle (0.1); 
\end{tikzpicture}\ .
\ee
The one-loop corrected propagator  is given by \rf{2.20},\rf{2.21}
(and is the  same in both  {\sf ZZ} and {\sf AdS} formulations). As a result, 
like in \rf{2.25},  the  contribution of each of the diagrams in \rf{3.7}  is given by 
the tree-level  expression  \rf{3.4} times 
 the  $1-\frac{1}{6} b^2$  factor, i.e.
\be\la{3.8}
C_{3,1}^{(1)} = C_{3,0}\times 3\times (-\tfrac{1}{6}) = \frac{8}{9}.
\ee

\paragraph{Diagrams with vertex tadpoles:}
In  the \ZZ formulation 
the contribution of  diagrams  with tadpoles attached to an internal 3-vertex 
\be
\la{3.11}
\begin{tikzpicture}[line width=1 pt, scale=0.4, rotate=0,baseline=-0.1cm]
\draw(-2,2)--(0,0); \draw(-2,0)--(0,0); \draw(-2,-2)--(0,0);
\begin{scope}[shift={(0,0)},scale=0.9]
	   \draw[fill=white,thin] (0,0) circle (0.3);
            \draw[thin] (45:0.3)--(225:0.3);
            \draw[thin] (-45:0.3)--(135:0.3);
	\end{scope}
\node[right] at (1,0) {$=$};
\end{tikzpicture}
\begin{tikzpicture}[line width=1 pt, scale=0.4, rotate=0,baseline=-0.1cm]
\draw(-2,2)--(0,0); \draw(-2,0)--(0,0); \draw(-2,-2)--(0,0); \draw (1,0) circle (1);
\node[right] at (2.5,0) {$+$};
\end{tikzpicture}
\begin{tikzpicture}[line width=1 pt, scale=0.4, rotate=0,baseline=-0.1cm]
\draw(-2,2)--(0,0); \draw(-2,0)--(0,0); \draw(-2,-2)--(0,0)--(1,0); \draw (2,0) circle (1);
\end{tikzpicture}
\ee
turns out to be   equivalent  
to that 
of the  tree diagram \rf{3.4}   with  the  cubic vertex 
$8\,b^{3}\,\frac{\chi^{3}}{3!}$ (the same as in \rf{31}  with $b\to b^3$)\foot{In  the \ZZ formulation 
$g(z,z)$ is  given by   \rf{2.14}   and  taking into account the  combinatorics (see \rf{A.8},(\ref{A.9}))
the contribution is  found to be the same as of  the tree diagram but with the coupling factor 
$-8b$ in \rf{3.4}  replaced by $-8\,b^{3}$. The same expression 
 appears in  Eq.~(15) of \cite{Menotti:2003km}.}
\be
C_{3,1}^{(2)}\,b^{3} = \ \ 
\begin{tikzpicture}[line width=1 pt, scale=0.4, rotate=0,baseline=-0.1cm]
\coordinate (A1) at (90:2);  \coordinate (A2) at (210:2);  \coordinate (A3) at (-30:2);
\coordinate (B1) at (90:1);  \coordinate (B2) at (210:1);  \coordinate (B3) at (-30:1);
\draw[densely dashed] (0,0) circle (2);
\draw (A1)--(0,0); \draw (A2)--(0,0); \draw (A3)--(0,0);
\draw[fill=black] (A1) circle (0.1); \draw[fill=black] (A2) circle (0.1); \draw[fill=black] (A3) circle (0.1); 
\begin{scope}[shift={(0,0)},scale=0.8]
	   \draw[fill=white,thin] (0,0) circle (0.3);
            \draw[thin] (45:0.3)--(225:0.3);
            \draw[thin] (-45:0.3)--(135:0.3);
	\end{scope}
\end{tikzpicture}\ . 
\ee
In the  {\sf AdS} formulation  we get the same  result (by a mechanism similar 
to the one in the case of  the two-point function \rf{2.20}--\rf{111}). 
Here 
 the tadpole  contributions   vanish ($g(z,z)=0$ in \rf{2.15}) 
but  the cubic vertex $\sim b\,\chi^{3}$ in \rf{31} is 
 rescaled   by $b\,Q = 1+b^{2}$  (see \rf{2.16}), i.e.  
 $-8b \chi^3  \to  -8b\chi^3 -8b^{3}\chi^3$    resulting    in  an additional one-loop correction. 
 Thus
\be\la{3.111}
C_{3,1}^{(2)} = C_{3,0} = -\frac{16}{9}.
\ee

\paragraph{Diagrams  with  cubic and quartic  vertices:}

There are also  one-loop  diagrams with one cubic and one quartic  vertices from \rf{31} 
\be
\la{3.13}
C_{3,1}^{(3)}\,b^{3} = \ \ 
\begin{tikzpicture}[line width=1 pt, scale=0.4, rotate=0,baseline=-0.1cm]
\coordinate (A1) at (90:2);  \coordinate (A2) at (230:2);  \coordinate (A3) at (-50:2);
\draw[densely dashed] (0,0) circle (2);
\draw (A1)--(0,0.7); \draw (0,0) circle(0.7); \draw (A2)--(0,-0.7)--(A3);
\draw[fill=black] (A1) circle (0.1); \draw[fill=black] (A2) circle (0.1); \draw[fill=black] (A3) circle (0.1); 
\node[right] at (2.8,0) {$+$ two  permutations \ . };
\end{tikzpicture}
\ee
Explicitly, we get (cf. \rf{3.4}) 
\be
\la{3.14}
\begin{tikzpicture}[line width=1 pt, scale=0.4, rotate=0,baseline=0]
\coordinate (A1) at (90:2);  \coordinate (A2) at (230:2);  \coordinate (A3) at (-50:2);
\draw[densely dashed] (0,0) circle (2);
\draw (A1)--(0,0.7); \draw (0,0) circle(0.7); \draw (A2)--(0,-0.7)--(A3);
\draw[fill=black] (A1) circle (0.1); \draw[fill=black] (A2) circle (0.1); \draw[fill=black] (A3) circle (0.1); 
\node[above] at (A1) {$\theta_{1}$};
\node[left] at (A2) {$\theta_{2}$};
\node[right] at (A3) {$\theta_{3}$};
\end{tikzpicture} \ \ = 
\frac{\frac{1}{2}\,(-8b)\,(-16b^{2})}{\mc K_{3}(\bm\theta)}\int \mathsf{d}^{2}z'\mathsf{d}^{2}z''\,
g_{\rm b}(\theta_{1}, z')\,g_{\rm b}(\theta_{2}, z'')\,g_{\rm b}(\theta_{3}, z'')\,\big[g(z', z'')\big]^{2},
\ee
where  $1\ov 2$ is the leg symmetry factor. 
Part  of this diagram (the loop with upper leg and integration over the cubic vertex point) is the same as   the function 
$\widetildeB(z_{2},z_{1})= \int\mathsf{d}^{2}z'\, \,g(z', z_{1})\, \big[g(z_{2}, z')\big]^{2}$ in \rf{A.11},(\ref{A.13}). 
Taking the point $z_1$ to the  boundary then gives,   according to (\ref{A.13}), 
the boundary-to-bulk propagator (\ref{2.27}) times 
the $1\ov 8$   factor  
\be
\lim_{|z_{1}|\to 1}\frac{1}{\mathsf{z}_{1}^{2}}\ \ 
\begin{tikzpicture}[line width=1 pt, scale=0.4, rotate=0,baseline=-0.1cm]
\draw (-2,0)--(-1,0); \draw (0,0) circle(1); 
\node[left] at (-1.8,0) {$z_{1}$};
\draw[fill=black] (1,0) circle(0.12); \draw[fill=black] (-2,0) circle(0.12);
\node[right] at (1.12,0) {$z_{2}$};
\end{tikzpicture}
= \frac{1}{8}
\begin{tikzpicture}[line width=1 pt, scale=0.4, rotate=0,baseline=-0.1cm]
\draw (-2,0)--(1,0);
\node[left] at (-1.8,0) {$\theta_{1}$};
\draw[fill=black] (1,0) circle(0.12); 
\node[right] at (1.12,0) {$z_{2}$};
\draw[thin] (200:2) arc(200:160:2);\draw[fill=black] (-2,0) circle(0.12);
\end{tikzpicture}
\ee
This  should  be  
multiplied by the  coupling constants from the two 
vertices  $(-8b)(-16 b^{2})$,  by  a symmetry factor $\frac{1}{2}$ and by  the diagram 
 multiplicity factor $3$.
 Compared to  the tree diagram contribution 
  this gives  an extra $-3\,b^{2}$  factor  in total, i.e. 
\be\la{3.15}
C_{3,1}^{(3)} = -3\,C_{3,0} = \frac{16}{3}\ .
\ee

\paragraph{Diagram with one-loop vertex correction:}
The most complicated  diagram 
 is the one with the three  $\chi^3$  vertices
and thus  with the three  bulk-to-bulk  and  the three bulk-to-boundary  propagators:
\be
\la{3.17}
C_{3,1}^{(4)}\,b^{3} = \ \ 
\begin{tikzpicture}[line width=1 pt, scale=0.4, rotate=0,baseline=-0.1cm]
\coordinate (A1) at (90:2);  \coordinate (A2) at (210:2);  \coordinate (A3) at (-30:2);
\coordinate (B1) at (90:1);  \coordinate (B2) at (210:1);  \coordinate (B3) at (-30:1);
\draw[densely dashed] (0,0) circle (2);
\draw (B1)--(B2)--(B3)--(B1);
\draw (A1)--(B1); \draw (A2)--(B2); \draw (A3)--(B3);
\draw[fill=black] (A1) circle (0.1); \draw[fill=black] (A2) circle (0.1); \draw[fill=black] (A3) circle (0.1); 
\end{tikzpicture}\ \ 
= \frac{(-8b)^{3}}{\mc K_{3}(\bm\theta)}\int \mathsf{d}^{2}z_{1}\mathsf{d}^{2}z_{2}\mathsf{d}^{2}z_{3}\,
g(z_{1},z_{2})\, g(z_{1},z_{3})\, g(z_{2},z_{3})\ 
\prod_{i=1}^{3}g_{\rm b}(\theta_{i}, z_{i})\ . 
\ee
As this is  the triple  \ads integral its analytic   computation is  non-trivial.  
According to \rf{3.6},\rf{3.66}   and the results in \rf{3.8},\rf{3.111},\rf{3.15}
the expected   value of $C_{3,1}^{(4)}$   should  be 
\be \la{316}
C_{3,1}^{(4)}= \frac{64}{27} -  \frac{8}{9}+\frac{16}{9}-\frac{16}{3}= -\frac{56}{27}\ . 
\ee
Since \rf{3.17}   should   give just  the value  of the   constant $C_{3,1}^{(4)}$
it is  sufficient to evaluate it numerically. 
We did this  by using the {\tt Suave} routine of the {\tt Cuba} library \cite{Hahn:2004fe}.
One important point is to check that  $C_{3,1}^{(4)}$  given by  \rf{3.17} is  indeed 
independent of  the boundary   coordinates 
$\theta_{1}, \theta_{2}, \theta_{3}$. 
From a practical perspective, one is interested in the $
\theta_{i}$ dependence of the systematic error in the numerical estimate at fixed computational effort (the total number of integrand evaluations).
\begin{figure}[htb]
\centering
\includegraphics[scale=0.4]{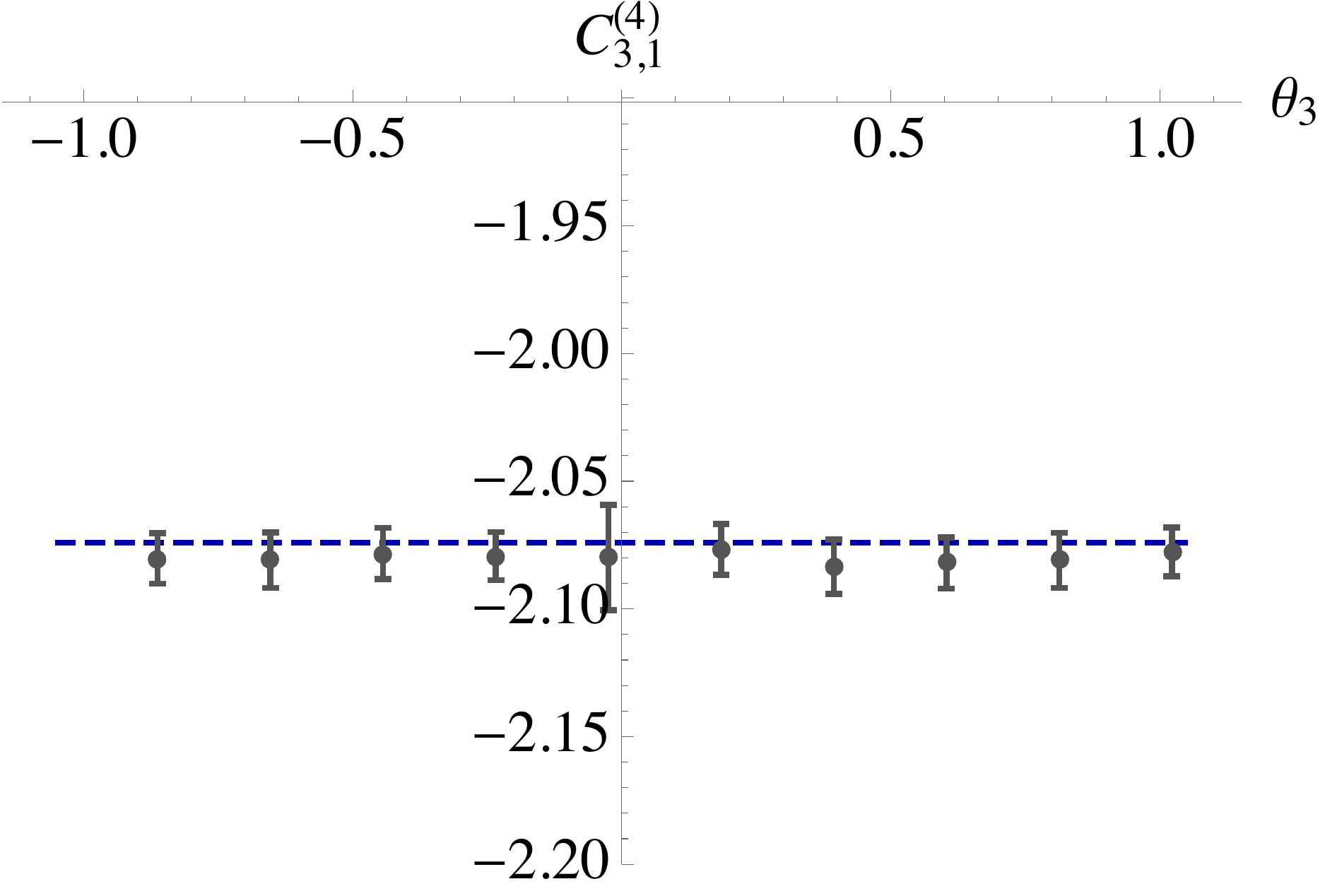}
\caption{Numerical estimate of $C^{(4)}_{3,1}$   extracted by $10^{7}$ 
integrand evaluations 
for $\theta_{1}=\frac{2\pi}{3}$, 
$\theta_{2}=-\frac{2\pi}{3}$ and variable $\theta_{3}$ with $|\theta_{3}|<\frac{\pi}{3}$. 
The horizontal blue line is the expected value $-\frac{56}{27}$.
}
\label{fig:num}
\end{figure}
  In Fig.~\ref{fig:num}
 we fixed the two points at
$\theta_{1}=\frac{2\pi}{3}$, $\theta_{2}=-\frac{2\pi}{3}$ and  varied 
 $\theta_{3}$ 
showing  the  resulting numerical estimate for $C_{3,1}^{(4)}$.
The horizontal line is the   expected value $-\frac{56}{27} $ 
 in \rf{316}.
One finds that  the integral  value  is more "flat" when the three external points $\theta_i$ 
 are well separated. This is expected as the  integrand  has a 
more singular behaviour  when at least two  external points get close. 
Considering  the equally spaced
points and increasing statistics we found the  best estimate to be\foot{The number in the 
  parenthesis  is  the statistical uncertainty (the least significant 
digits of a given numerical value for the average). Here the uncertainty is the  standard deviation 
of the Monte-Carlo evaluation of the loop integral.  Explicitly,   
$ 2.08(2)\equiv 2.08\pm 0.02$.}
\be
\la{3.18}
C_{3,1}^{(4)}  = -2.08(2) \ .
\ee
This  is in good agreement with \rf{316} or  
 $-\frac{56}{27} \approx  -2.0741$.
 
 The expected  value \rf{3.6}   and thus \rf{316}  was found 
 from \rf{1.9}   which was   based on matching the two-point and three-point  correlators \rf{1.2} 
 with the stress tensor ones  \rf{1.4}  for the Liouville   central  charge  value 
 $c=1 + 6 ( b + b^{-1})^2  $. 
 Alternatively, we may   determine the   values of $c$    and $\k$ in \rf{1.7} 
 from the   explicit  tree  plus one-loop  results for $C_2$ and $C_3$  found above. 
Omitting the overall  conformally covariant factors 
we may symbolically summarize the results for the two-point and three-point correlators in 
\rf{2.25}    and   \rf{2.29},\rf{3.8},\rf{3.11},\rf{3.15},\rf{3.18}    as follows 
\begin{align}\la{3.20}
\langle \Phi\Phi\rangle = \tfrac{4}{3}\,(1-\tfrac{1}{6}\,b^{2})+\mc O(b^{4}), \qquad \qquad 
\langle \Phi\Phi\Phi\rangle = -\tfrac{16}{9}\,b+\big(\tfrac{8}{9}-\tfrac{16}{9}+\tfrac{16}{3}+C_{3,1}^{(4)}\big)\,b^{3}
+\mc O(b^{5}).
\end{align}
The consistency of \rf{3.20} 
 with the identification in \rf{1.3},\rf{1.4} or \rf{1.7} 
\be\la{3.21}
\langle \Phi\Phi\rangle = \ha c\, \kappa^{2}, \qquad\qquad 
\langle \Phi\Phi\Phi\rangle = c\,  \kappa^{3}\ 
\ee
implies that
\begin{align}
\la{3.22}
\kappa = -\tfrac{2}{3}\,b+\big(\tfrac{14}{9}+\tfrac{3}{8}\,C_{3,1}^{(4)}\big)\,b^{3}+\mc O(b^{5}),\qquad \qquad 
c = 6\, {b^{-2}}+27\,(1+\tfrac{1}{4}\,C_{3,1}^{(4)})+\mc O(b^{2}).
\end{align}
Using the numerical estimate  for  $C_{3,1}^{(4)}$  found in  (\ref{3.18}) gives 
\be\la{3.23}
c = 6\,{b^{-2}}+13.0(1)+\mc O(b^{2}),
\ee
which is  in good  agreement with 
 the  value of the Liouville  central charge     $c = 6\,{b^{-2}}+13+ 6\, b^{2}$. 
 Note  also that  for  $C_{3,1}^{(4)} = -\frac{56}{27}$  corresponding to the exact value of $c$
 we get 
\begin{align}
\la{3.24}
\kappa = -\frac{2}{3}\,b\,\big(1-\frac{7}{6}\,b^{2}\big)+\mc O(b^{5})\ , 
\end{align}
and thus the following expressions for $C_2$ \rf{2.25} and $C_3$ in \rf{1.6},\rf{1.8}
\be \la{3.259}
C_2=  \frac{4}{3}\big(1-\frac{1}{6}\,b^{2}\big)+\mc O(b^{4})   \ , \qquad \qquad 
C_3 = - \frac{16}{9} b\big(1-\frac{4}{3}\,b^{2}\big)+\mc O(b^{5})   \ .
\ee

\section{Four-point boundary correlation function}
\la{sec:fourpoints}

The four-point function \rf{1.11}  receives  contributions from the 
 disconnected and connected  diagrams in AdS$_{2}$.  
The disconnected part is given by 
 $\langle \Phi(t_{1})\, \Phi(t_{2})\rangle\,
\langle\Phi(t_{3})\, \Phi(t_{4})\rangle\ ($+ permutations) 
  corresponding to \rf{1.12}. 
 From (\ref{3.21}),\rf{3.25} the prediction for the 
non-trivial  connected part is 
\begin{align}
\la{3.25}
\langle \Phi(t_{1})\, \Phi(t_{2})\, 
\Phi(t_{3})\, \Phi(t_{4}) \rangle_{\rm conn} = &C_{4}(b)\,\Big(\frac{1}{t_{12}^{2}\,
t_{23}^{2}\,t_{34}^{2}\,t_{14}^{2}}
+\frac{1}{t_{13}^{2}\,t_{24}^{2}\,t_{14}^{2}\,t_{23}^{2}}
+\frac{1}{t_{12}^{2}\,t_{24}^{2}\,t_{34}^{2}\,t_{13}^{2}}\Big),\\
\la{3.26}
C_{4}(b) =&c\,  \kappa^{4} = \tfrac{32}{27}\,b^{2}-\tfrac{80}{27}\,b^{4}+\mc O(b^{6}),
\end{align}
implying that in \rf{1.14} one should have 
\be \la{4.1}
C_{4,0}= \tfrac{32}{27}, \qquad \qquad  C_{4,1}= -\tfrac{80}{27} \ . \ee
These values are in  agreement with  the  relations in  (\ref{1.15})
 for the coefficients 
$C_{2,0}=\frac{4}{3}$, $C_{2,1}=-\frac{2}{9}$, $C_{3,0}=-\frac{16}{9}$, $C_{3,1}=\frac{64}{27}$  found above. 

The   value of  the leading coefficient  $C_{4,0}$    follows  from 
the sum of   the   tree-level   contact diagram   and the  the three exchange diagrams as in  \rf{1.10}  (see Appendix \ref{app:tree})
\be\la{4.4}
C_{4,0}\,b^{2} = ( C_{4,0}^{\rm cont}  +  C_{4,0}^{\rm exch}) \, b^2 =   \ \ 
\begin{tikzpicture}[line width=1 pt, scale=0.4, rotate=0,baseline=-0.1cm]
\coordinate (A1) at (135:2);  \coordinate (A2) at (45:2);  
\coordinate (A3) at (-45:2);   \coordinate (A4) at (-135:2);
\draw[densely dashed] (0,0) circle (2);
\draw (A1)--(0,0)--(A2); \draw (A3)--(0,0)--(A4);
\draw[fill=black] (A1) circle (0.1); \draw[fill=black] (A2) circle (0.1); 
\draw[fill=black] (A3) circle (0.1); \draw[fill=black] (A4) circle (0.1); 
\end{tikzpicture}\ \ + \bigg[\, 
\begin{tikzpicture}[line width=1 pt, scale=0.4, rotate=0,baseline=-0.1cm]
\coordinate (A1) at (135:2);  \coordinate (A2) at (45:2);  
\coordinate (A3) at (-45:2);   \coordinate (A4) at (-135:2);
\coordinate (B1) at (-1,0); \coordinate (B2) at (1,0);
\draw[densely dashed] (0,0) circle (2);
\draw (A1)--(B1)--(A4); \draw (A3)--(B2)--(A2); \draw (B1)--(B2);
\draw[fill=black] (A1) circle (0.1); \draw[fill=black] (A2) circle (0.1); 
\draw[fill=black] (A3) circle (0.1); \draw[fill=black] (A4) circle (0.1); 
\end{tikzpicture}\ \ +\text{crossed}\bigg] \ \
= \tfrac{32}{27}\,b^{2}.
\ee
 The  one-loop coefficient $C_{4,1} $  is determined   from 
  the sum of the contributions of the   five classes of diagrams
\be 
C_{4,1} =C^{(0)}_{4,1}   + \sum_{i=1}^4   C^{(i)}_{4,1} \ . \la{4.2}\ee
To   compute  them  we shall use the AdS  formulation  in which  there are  no tadpoles
but the potential   has an extra factor $(1+ b^2)$ in \rf{2.16}  rescaling the
 vertices. 
$C^{(0)}_{4,1} $ will denote   the resulting contribution  originating, due to this rescaling, 
from the  contact   and exchange 
tree diagrams in \rf{4.4}.  Since  the contact  4-vertex is rescaled by $1+b^{2}$ while the exchange diagrams with  two 3-vertices 
by $(1+b^{2})^{2}$  the two  are  combined  now with the  relative factor of 2 
 (cf. \rf{4.4})
\be\la{4.6}
C_{4,1}^{(0)} = C_{4,0}^{\rm cont}+2\,C_{4,0}^{\rm exch}\ .
\ee
The  contributions $C^{(i)}_{4,1}$  with $i=1,2,3,4$ in  \rf{4.2}  come from the  genuine one-loop diagrams
described  below. 

In writing \rf{4.2} and \rf{4.6}  we formally assumed  that $t_i$-dependence 
of  each  individual  diagram contribution 
is given by  the same  simple  conformal factor  as in \rf{3.25}  (cf.  \rf{3.3},\rf{2.26})
\be\la{4.9}
\mc K_{4}(\bm\theta)\equiv \mc K_{4}(\theta_{1}, \dots, \theta_{4}) = 
\frac{1}{t_{12}^{2}\,
t_{23}^{2}\,t_{34}^{2}\,t_{14}^{2}}
+\frac{1}{t_{13}^{2}\,t_{24}^{2}\,t_{14}^{2}\,t_{23}^{2}}
+\frac{1}{t_{12}^{2}\,t_{24}^{2}\,t_{34}^{2}\,t_{13}^{2}}, \quad 
\ \ t_{ij} = t(\theta_{i})-t(\theta_{j}),
\ee
However, this   need not be true in general.  Let us first   recall  that 
  the  tree level  four-point  boundary correlator \rf{1.10} 
 has the canonical form in \rf{1.12},\rf{1.13} 
 without any  logarithmic terms depending on the 1d  invariant cross-ratio
  $\chi={t_{12} t_{34} \ov t_{13} t_{24}}
$. Such terms  cancel
between  the contact and exchange diagram  contributions     \cite{Ouyang:2019xdd}
(see Appendix \ref{app:tree})  but 
 they   survive  if  the  contact and exchange contributions are combined 
with a different relative coefficient like in \rf{4.6}  (see (\ref{D.2})).
These  extra terms  should  still not appear in the total expression \rf{3.25} 
for the one-loop corrected  correlator, i.e. they   should  cancel  against   similar "extra" 
 contributions of   other diagrams. Checking  this  directly  would require the
  analytic computation of all  the  diagrams   discussed below  which we will not attempt here. 
  
 Here we will  compute  the total  value of the coefficient $C_{4,1}$ numerically
 by fixing particular values of the boundary coordinates $t_i$.
 We will   also give the  values of the individual diagram contributions  $C^{(k)}_{4,1}$
 defined formally as the corresponding  integrals   divided   by the factor $\mc K_{4}$ in \rf{4.9}.

\iffa
From a practical perspective, 
we shall assume  that  the sum of all one-loop diagrams contributing to  the four-point 
 boundary correlator  or to $C_{4,1}$ in \rf{4.2} 
 will   have the simple   conformal structure  proportional to the factor (cf.  \rf{3.3},\rf{2.26})
as in \rf{3.25}. This   need not be true   for the  contributions  of the 
 individual diagrams.\footnote{Let us note,  We should  be able still  to  evaluate  the individual diagram contributions  numerically 
 by simply assuming that  all the additional  contributions   not  proportional to 
 $\mc K_{4}(\bm\theta)$      cancel in the sum. 
 After this preliminary remark, let us 
 consider the various classes of one-loop corrections, i.e. the terms with $i=1, \dots, 4$ in (\ref{4.2}).
\footnote{Let us note 
 that  the  tree level  four-point  boundary correlator \rf{1.10} 
 has the canonical form in \rf{1.12},\rf{1.13} 
 without any  logarithmic terms depending on the 1d  invariant cross-ratio
  $\chi={t_{12} t_{34} \ov t_{13} t_{24}}
$:    such terms  cancel
between  the contact and exchange diagram  contributions     \cite{Ouyang:2019xdd}. 
However, they may survive  in the  contact and exchange contributions are combined 
with a different relative coefficient like in \rf{4.6}. 
Such terms should should still not appear in the total expression 
for the one-loop corrected  correlator   so they  should be cancelled  by 
a  contribution coming from some other diagrams.
Checking  this  directly  would require the analytic computation of all  the  diagrams   discussed below  which we will not attempt here. 
 }
 \fi 

\paragraph{Diagrams with dressed  external propagators:} 
These diagrams are analogous to the ones in \rf{3.7}  (their contribution is actually 
proportional to $K_{4}(\bm\theta)$). 
As in \rf{3.7},\rf{3.8}   the  one-loop   dressing of  the bulk-to-boundary propagators
gives the contribution 
\be\la{4.66}
C_{4,1}^{(1)} = C_{4,0}\times 4 \times (-\tfrac{1}{6}) = -\tfrac{64}{81}.
\ee
\paragraph{Diagrams with dressed internal propagator:}
\be
\la{3.29}
C_{4,1}^{(2)}\,b^{4} = 
\begin{tikzpicture}[line width=1 pt, scale=0.4, rotate=0,baseline=-0.1cm]
\coordinate (A1) at (135:2);  \coordinate (A2) at (45:2);  
\coordinate (A3) at (-45:2);   \coordinate (A4) at (-135:2);
\coordinate (B1) at (-1,0); \coordinate (B2) at (1,0);
\draw[densely dashed] (0,0) circle (2);
\draw (A1)--(B1)--(A4); \draw (A3)--(B2)--(A2); \draw (B1)--(B2);
\draw[fill=lightgray] (0,0) circle(0.5);
\draw[fill=black] (A1) circle (0.1); \draw[fill=black] (A2) circle (0.1); 
\draw[fill=black] (A3) circle (0.1); \draw[fill=black] (A4) circle (0.1); 
\end{tikzpicture}\ \ +\text{crossed}\ . 
\ee
Here the grey circle stands for the one-loop self energy correction 
$\Sigma(z,z')$ in  (\ref{2.21}) or explicitly
\be
\begin{tikzpicture}[line width=1 pt, scale=0.4, rotate=0,baseline=-0.1cm]
\coordinate (A1) at (135:2);  \coordinate (A2) at (45:2);  
\coordinate (A3) at (-45:2);   \coordinate (A4) at (-135:2);
\coordinate (B1) at (-1,0); \coordinate (B2) at (1,0);
\draw[densely dashed] (0,0) circle (2);
\draw (A1)--(B1)--(A4); \draw (A3)--(B2)--(A2); \draw (B1)--(B2);
\draw[fill=lightgray] (0,0) circle(0.5);
\draw[fill=black] (A1) circle (0.1); \draw[fill=black] (A2) circle (0.1); 
\draw[fill=black] (A3) circle (0.1); \draw[fill=black] (A4) circle (0.1); 
\node[above] at (A1) {$\theta_{1}$}; \node[above] at (A2) {$\theta_{2}$};
\node[below] at (A3) {$\theta_{3}$}; \node[below] at (A4) {$\theta_{4}$};
\end{tikzpicture}\ \ = \frac{(-8b)^{2}}{\mc K_{4}(\bm\theta)}\,\int\mathsf{d}^{2}z'
\mathsf{d}^{2}z''\,\Sigma(z',z'')
\,g_{\rm b}(\theta_{1}, z')\,g_{\rm b}(\theta_{4}, z')
\,g_{\rm b}(\theta_{2}, z'')\,g_{\rm b}(\theta_{3}, z''). \la{4.8}
\ee
\paragraph{Diagrams with one-loop 1PI cubic vertex correction:}
\be
C_{4,1}^{(3)}\,b^{4} = 
\begin{tikzpicture}[line width=1 pt, scale=0.4, rotate=0,baseline=-0.1cm]
\coordinate (A1) at (135:2);  \coordinate (A2) at (45:2);  
\coordinate (A3) at (-45:2);   \coordinate (A4) at (-135:2);
\coordinate (B1) at (-1,0); \coordinate (B2) at (1,0);
\draw[densely dashed] (0,0) circle (2);
\draw (A1)--(B1)--(A4); \draw (A3)--(B2)--(A2); \draw (B1)--(B2);
\draw[fill=lightgray] (B1) circle(0.5);
\draw[fill=black] (A1) circle (0.1); \draw[fill=black] (A2) circle (0.1); 
\draw[fill=black] (A3) circle (0.1); \draw[fill=black] (A4) circle (0.1); 
\end{tikzpicture}\ \ +
\begin{tikzpicture}[line width=1 pt, scale=0.4, rotate=0,baseline=-0.1cm]
\coordinate (A1) at (135:2);  \coordinate (A2) at (45:2);  
\coordinate (A3) at (-45:2);   \coordinate (A4) at (-135:2);
\coordinate (B1) at (-1,0); \coordinate (B2) at (1,0);
\draw[densely dashed] (0,0) circle (2);
\draw (A1)--(B1)--(A4); \draw (A3)--(B2)--(A2); \draw (B1)--(B2);
\draw[fill=lightgray] (B2) circle(0.5);
\draw[fill=black] (A1) circle (0.1); \draw[fill=black] (A2) circle (0.1); 
\draw[fill=black] (A3) circle (0.1); \draw[fill=black] (A4) circle (0.1); 
\end{tikzpicture}
\ \ + \text{crossed}\ . \la{4.10}
\ee
Here the grey circle stands for the one-loop cubic vertex correction.
In the  {\sf AdS} formulation where the tadpoles are absent  it is given by the 1PI 
diagrams 
\be\la{4.11} 
\begin{tikzpicture}[line width=1 pt, scale=0.4, rotate=0,baseline=-0.1cm]
\coordinate (A1) at (90:2);  \coordinate (A2) at (210:2);  \coordinate (A3) at (-30:2);
\coordinate (B1) at (90:1);  \coordinate (B2) at (210:1);  \coordinate (B3) at (-30:1);
\draw (A1)--(0,0); \draw (A2)--(0,0); \draw (A3)--(0,0);
\draw[fill=lightgray] (0,0) circle(0.6);
\draw[fill=black] (A1) circle (0.1); \draw[fill=black] (A2) circle (0.1); \draw[fill=black] (A3) circle (0.1); 
\end{tikzpicture}\ \ = 
\begin{tikzpicture}[line width=1 pt, scale=0.4, rotate=0,baseline=-0.1cm]
\coordinate (A1) at (90:2);  \coordinate (A2) at (210:2);  \coordinate (A3) at (-30:2);
\coordinate (B1) at (90:1);  \coordinate (B2) at (210:1);  \coordinate (B3) at (-30:1);
\draw (B1)--(B2)--(B3)--(B1);
\draw (A1)--(B1); \draw (A2)--(B2); \draw (A3)--(B3);
\draw[fill=black] (A1) circle (0.1); \draw[fill=black] (A2) circle (0.1); \draw[fill=black] (A3) circle (0.1); 
\end{tikzpicture}\ \ + \bigg[
\begin{tikzpicture}[line width=1 pt, scale=0.4, rotate=0,baseline=-0.1cm]
\coordinate (A1) at (90:2);  \coordinate (A2) at (230:2);  \coordinate (A3) at (-50:2);
\draw (A1)--(0,0.7); \draw (0,0) circle(0.7); \draw (A2)--(0,-0.7)--(A3);
\draw[fill=black] (A1) circle (0.1); \draw[fill=black] (A2) circle (0.1); \draw[fill=black] (A3) circle (0.1); 
\end{tikzpicture}\ \ + \text{crossed}\ 
\bigg] \ . 
\ee
 The
expressions  for the resulting integrals are similar  to  the one-loop three-point 
contributions  in (\ref{3.14}) and (\ref{3.17})  connected to the  remaining  tree parts.

\paragraph{Diagrams   with 1PI quartic vertex corrections:}
The final  class of diagrams represent  irreducible quartic vertex
corrections (here we do not indicate explicitly additional diagrams obtained by 
exchanging the boundary points)
\be\la{4.13}
C_{4,1}^{(4)}\,b^{4} = 
\begin{tikzpicture}[line width=1 pt, scale=0.4, rotate=0,baseline=-0.1cm]
\coordinate (A1) at (135:2);  \coordinate (A2) at (45:2);  
\coordinate (A3) at (-45:2);   \coordinate (A4) at (-135:2);
\draw[densely dashed] (0,0) circle (2);
\draw (A1)--(0,0)--(A2); \draw (A3)--(0,0)--(A4);
\draw[fill=white] (0,0) circle(0.7);
\draw[fill=black] (A1) circle (0.1); \draw[fill=black] (A2) circle (0.1); 
\draw[fill=black] (A3) circle (0.1); \draw[fill=black] (A4) circle (0.1); 
\node at (0,-2.8) {(a)};
\end{tikzpicture}\ +
\begin{tikzpicture}[line width=1 pt, scale=0.4, rotate=0,baseline=-0.1cm]
\coordinate (A1) at (135:2);  \coordinate (A2) at (45:2);  
\coordinate (A3) at (-45:2);   \coordinate (A4) at (-135:2);
\draw[densely dashed] (0,0) circle (2);
\draw (A1)--(0,0)--(A2); \draw (A3)--(0,-0.7)--(A4);
\draw[fill=white] (0,0) circle(0.7);
\draw[fill=black] (A1) circle (0.1); \draw[fill=black] (A2) circle (0.1); 
\draw[fill=black] (A3) circle (0.1); \draw[fill=black] (A4) circle (0.1); 
\node at (0,-2.8) {(b)};
\end{tikzpicture}\ +
\begin{tikzpicture}[line width=1 pt, scale=0.4, rotate=0,baseline=-0.1cm]
\coordinate (A1) at (135:2);  \coordinate (A2) at (45:2);  
\coordinate (A3) at (-45:2);   \coordinate (A4) at (-135:2);
\draw[densely dashed] (0,0) circle (2);
\draw (A1)--(0,0.7)--(A2); \draw (A3)--(0,-0.7)--(A4);
\draw[fill=white] (0,0) circle(0.7);
\draw[fill=black] (A1) circle (0.1); \draw[fill=black] (A2) circle (0.1); 
\draw[fill=black] (A3) circle (0.1); \draw[fill=black] (A4) circle (0.1); 
\node at (0,-2.8) {(c)};
\end{tikzpicture}\ +
\begin{tikzpicture}[line width=1 pt, scale=0.4, rotate=0,baseline=-0.1cm]
\coordinate (A1) at (90:2);  \coordinate (A2) at (-45:2);  
\coordinate (A3) at (-90:2);   \coordinate (A4) at (-135:2);
\draw[densely dashed] (0,0) circle (2);
\draw (A1)--(0,0); \draw (0,-0.7)--(A2); \draw (A3)--(0,-0.7)--(A4);
\draw[fill=white] (0,0) circle(0.7);
\draw[fill=black] (A1) circle (0.1); \draw[fill=black] (A2) circle (0.1); 
\draw[fill=black] (A3) circle (0.1); \draw[fill=black] (A4) circle (0.1); 
\node at (0,-2.8) {(d)};
\end{tikzpicture}  
\ee
The  integrals  corresponding to   these   diagrams are 
\begin{align}
\begin{tikzpicture}[line width=1 pt, scale=0.4, rotate=0,baseline=-0.1cm]
\coordinate (A1) at (135:2);  \coordinate (A2) at (45:2);  
\coordinate (A3) at (-45:2);   \coordinate (A4) at (-135:2);
\draw[densely dashed] (0,0) circle (2);
\draw (A1)--(0,0)--(A2); \draw (A3)--(0,0)--(A4);
\draw[fill=white] (0,0) circle(0.7);
\draw[fill=black] (A1) circle (0.1); \draw[fill=black] (A2) circle (0.1); 
\draw[fill=black] (A3) circle (0.1); \draw[fill=black] (A4) circle (0.1); 
\node[above] at (A1) {$\theta_{1}$}; \node[above] at (A2) {$\theta_{2}$};
\node[below] at (A3) {$\theta_{3}$}; \node[below] at (A4) {$\theta_{4}$};
\end{tikzpicture}\ \ &= \frac{(-8b)^{4}}{\mc K_{4}(\bm\theta)}\,\int\prod_{i=1}^{4}\mathsf{d}^{2}z_{i}
\prod_{i=1}^{4}g_{\rm b}(\theta_{i}, z_{i})\,\, g(z_{1},z_{2})\, g(z_{2},z_{3})\, g(z_{3},z_{4})\, g(z_{1},z_{4}),  \no\\
\begin{tikzpicture}[line width=1 pt, scale=0.4, rotate=0,baseline=-0.1cm]
\coordinate (A1) at (135:2);  \coordinate (A2) at (45:2);  
\coordinate (A3) at (-45:2);   \coordinate (A4) at (-135:2);
\draw[densely dashed] (0,0) circle (2);
\draw (A1)--(0,0)--(A2); \draw (A3)--(0,-0.7)--(A4);
\draw[fill=white] (0,0) circle(0.7);
\draw[fill=black] (A1) circle (0.1); \draw[fill=black] (A2) circle (0.1); 
\draw[fill=black] (A3) circle (0.1); \draw[fill=black] (A4) circle (0.1); 
\node[above] at (A1) {$\theta_{1}$}; \node[above] at (A2) {$\theta_{2}$};
\node[below] at (A3) {$\theta_{3}$}; \node[below] at (A4) {$\theta_{4}$};
\end{tikzpicture} \ \ &= 
\frac{(-8b)^{2}(-16b^{2})}{\mc K_{4}(\bm\theta)}\,\int \mathsf{d}^{2}z'\mathsf{d}^{2}z''\mathsf{d}^{2}z'''
g_{\rm b}(\theta_{1}, z')\,g_{\rm b}(\theta_{2}, z'')\,g_{\rm b}(\theta_{3}, z''')\,
g_{\rm b}(\theta_{4}, z''')\notag \\
& \qquad \qquad \qquad \qquad\qquad \qquad \ \ \times  g(z',z'')\ g(z',z''')\ g(z'',z'''),  \la{4.14} \\
\begin{tikzpicture}[line width=1 pt, scale=0.4, rotate=0,baseline=-0.1cm]
\coordinate (A1) at (135:2);  \coordinate (A2) at (45:2);  
\coordinate (A3) at (-45:2);   \coordinate (A4) at (-135:2);
\draw[densely dashed] (0,0) circle (2);
\draw (A1)--(0,0.7)--(A2); \draw (A3)--(0,-0.7)--(A4);
\draw[fill=white] (0,0) circle(0.7);
\draw[fill=black] (A1) circle (0.1); \draw[fill=black] (A2) circle (0.1); 
\draw[fill=black] (A3) circle (0.1); \draw[fill=black] (A4) circle (0.1); 
\node[above] at (A1) {$\theta_{1}$}; \node[above] at (A2) {$\theta_{2}$};
\node[below] at (A3) {$\theta_{3}$}; \node[below] at (A4) {$\theta_{4}$};
\end{tikzpicture}\ \ &= 
\frac{\frac{1}{2}(-16b^{2})^{2}}{\mc K_{4}(\bm\theta)}\,\int \mathsf{d}^{2}z'\mathsf{d}^{2}z''\,
g_{\rm b}(\theta_{1}, z')\,g_{\rm b}(\theta_{2}, z')\,g_{\rm b}(\theta_{3}, z'')\,
g_{\rm b}(\theta_{4}, z'')\,\big[g(z',z'')\big]^{2}, \notag \\
\begin{tikzpicture}[line width=1 pt, scale=0.4, rotate=0,baseline=-0.1cm]
\coordinate (A1) at (90:2);  \coordinate (A2) at (-45:2);  
\coordinate (A3) at (-90:2);   \coordinate (A4) at (-135:2);
\draw[densely dashed] (0,0) circle (2);
\draw (A1)--(0,0); \draw (0,-0.7)--(A2); \draw (A3)--(0,-0.7)--(A4);
\draw[fill=white] (0,0) circle(0.7);
\draw[fill=black] (A1) circle (0.1); \draw[fill=black] (A2) circle (0.1); 
\draw[fill=black] (A3) circle (0.1); \draw[fill=black] (A4) circle (0.1); 
\node[above] at (A1) {$\theta_{1}$}; \node[below] at (A2) {$\theta_{2}$};
\node[below] at (A3) {$\theta_{3}$}; \node[below] at (A4) {$\theta_{4}$};
\end{tikzpicture} \ \ &= 
\frac{\frac{1}{2}(-8b)(-32b^{3})}{\mc K_{4}(\bm\theta)}\,\int \mathsf{d}^{2}z'\mathsf{d}^{2}z''\,
g_{\rm b}(\theta_{1}, z')\,g_{\rm b}(\theta_{2}, z'')\,g_{\rm b}(\theta_{3}, z'')\,
g_{\rm b}(\theta_{4}, z'')\,\big[g(z',z'')\big]^{2}.\notag
\end{align}
\iffa 
The total one-loop coefficient is 
\be
C_{4,1} = -\tfrac{64}{81}+C_{4,1}^{(\sf AdS)}+
C_{4,1}^{(2)}+C_{4,1}^{(3)}+C_{4,1}^{(4)}.
\ee
\fi
 
 We evaluated $C^{(i)}_{4,1}$ by the numerical integration 
for the  boundary point  configurations where  the four  
$\theta_{i}$ are almost equally spaced. 
As was  discussed
above  in the case of the three-point function, 
such  choice is expected to minimize the systematic numerical error. 
The integrator routine is again chosen to be {\tt Suave}. 
The most challenging  is the computation  of  $C_{4,1}^{(3)}$  in \rf{4.10} 
for which we evaluated the integrand
at about $10^{8}$ points. 
To give  an idea of the relative weight of the 
different  contributions, we quote them 
for the symmetric configuration $\bm\theta = (\frac{\pi}{4}, \frac{3\pi}{4}, \frac{5\pi}{4}, \frac{7\pi}{4})$ 
\be  \def\arraystretch{1.3}
\begin{array}{cccccc}
\toprule
 C_{4,1}^{(0)} &  C_{4,1}^{(1)}  & C_{4,1}^{(2)} & C_{4,1}^{(3)} & C_{4,1}^{(4)} &\ \  C_{4,1}^{(\rm tot)} 
  \\
\midrule
2.675(3) & -\frac{64}{81}=-0.7901...  &  -0.2191(2) & -5.54(2) &  0.884(6) &\ \  -2.99(3)\\
\bottomrule
\end{array}
\la{4.15}
\ee
The resulting  total value $C^{(\rm tot)} _{4,1}= -2.99\pm 0.03 $ 
given by the sum in \rf{4.2} is  to be compared with the expected value in 
(\ref{3.26}), i.e.   $C_{4,1} = -\frac{80}{27} = -2.963...$.
The relative error  is thus  below the  1\%  level and within the statistical uncertainty of the multidimensional integration.
This very good agreement with the prediction \rf{4.1} 
effectively confirms  our assumption  that  like the tree-level one,  the 
total one-loop correlator   has simple  dependence on $t_i$  as in \rf{3.25}.

\section*{Acknowledgments}

We are  grateful to Simone Giombi,  Erik Tonni   and Xi Yin  for  very useful discussions.
AAT was supported by the STFC grant ST/P000762/1.
AAT also thanks the Galileo Galilei Institute for Theoretical Physics for hospitality  and  INFN for partial support during the completion of this work. 

\appendix

\section{\ads integrals in  disc parametrization}
\la{app:disc}

Below we  collect  some results about 
the $SU(1,1)$ covariant  integrals over the unit disc 
that were used  in the main text. 
We shall use the  invariant measure 
\be
\la{A.1}
\mathsf{d}^{2}z = \frac{d^{2}z}{\pi\,(1-|z|^{2})^{2}}, 
\ee
and  the propagator $g(z,z')$ defined in (\ref{2.5}) with $g(z,z)$ parametrized  as in 
(\ref{2.13}). 
A very useful representation of the propagator is the Fourier expansion derived in \cite{Menotti:2003km}.
Denoting by  $\vt(z,z')$  the angle between the disc points $z,z'$, one has 
\begin{align}
\la{A.2}
& \qquad  g(z,z') = \sum_{n=0}^{\infty}g_{n}(|z|^{2},|z'|^{2})\,\cos(n\,\vt(z,z')), \\
g_{n}(x,y) &= \theta(y-x)\,a_{n}(x)\,b_{n}(y)+\theta(x-y)\,a_{n}(y)\,b_{n}(x),\la{a3} \\
a_{0}(x) &= \frac{1+x}{1-x},\ \ \qquad  \qquad  \qquad  \quad \qquad
a_{1}(x) = \frac{\sqrt x}{1-x},\no  \\
 b_{0}(y) &= -\frac{1}{2}\,\Big(\frac{1+y}{1-y}\log y+2\Big), \ \ \qquad  \ \   b_{1}(y) = \frac{1}{\sqrt y}\,\Big(\frac{2y}{1-y}\,\log y+1+y\Big), \notag \\
a_{n\ge 2}(x) &= \frac{x^{\frac{n}{2}}}{1-x}\,\Big(1-\frac{n-1}{n+1}\,x\Big),  \qquad \ \ 
b_{n\ge 2}(y) = -\frac{y^{-\frac{n}{2}}}{n\,(n-1)}\,\Big(\frac{1+y}{1-y}\,(1-y^{n})
-n\,(1+y^{n})\Big) ,
\end{align}
where $\theta(y-x)$ in \rf{a3}   is the  step function.

\subsection{Tadpole  reduction identity}

Let us  consider the integral  (using (\ref{2.13}))
\begin{align}
\la{A.4}
I(z) = \int\mathsf{d}^{2}z'\,g(z,z')\,g(z',z') = \int\mathsf{d}^{2}z'\,g(z,z')\,
\big[\rC_{1}\log(1-|z'|^{2})-\rC_{2}\big]\ . 
\end{align}
Substituting   (\ref{A.2})
the only  contributing  term is $n=0$. After integrating  over the angle and setting $|z'|=r$ we have 
\begin{align}
I(z) = &2\,\pi\,\int_{0}^{1}\frac{r\,dr}{\pi\,(1-r^{2})^{2}}\,g_{0}(|z|^{2},r^{2})\,
\big[\C_{1}\log(1-r^{2})-\C_{2}\big]\notag\\
=& 2\,\frac{1+|z|^{2}}{1-|z|^{2}}\,\int_{|z|}^{1}\frac{r\,dr}{(1-r^{2})^{2}}\,b_{0}(r^{2})\,
\big[\C_{1}\log(1-r^{2})-\C_{2}\big]\no\\
&+2\,b_{0}(|z|^{2})\,\int_{0}^{|z|}\frac{r\,dr}{(1-r^{2})^{2}}\,\frac{1+r^{2}}{1-r^{2}}\,
\big[\C_{1}\log(1-r^{2})-\C_{2}\big]\ . 
\end{align}
Integrating over $x=r^2$   gives 
\be
\la{A.6}
I(z) = -\frac{1}{8} (\C_{1}+2\,\C_{2}) +\frac{1}{4} \C_{1}\,\log(1-|z|^{2}).
\ee
Combining \rf{A.4}  and \rf{A.6} we get  a simple relation (which 
 follows also from the defining equation for  the 
propagator)
\be
\la{A.7}
g(z,z)-4\,\int\mathsf{d}^{2}z'\,g(z,z')\,g(z',z') = \frac{\C_{1}}{2} \ . 
\ee
It  may be used to perform  the  following reduction of the  tadpole diagrams  
(special cases  of   which are mentioned in \cite{Menotti:2003km}) 
\be
\la{A.8}
\begin{tikzpicture}[line width=1 pt, scale=0.4, rotate=0,baseline=0]
\draw(-2,2)--(0,0); \draw(-2,-2)--(0,0); \draw (1,0) circle (1);
 \draw[fill=black] (-1.5,0.9) circle(0.03); 
 \draw[fill=black] (-1.6,0.5) circle(0.03); 
 \draw[fill=black] (-1.65,0.0) circle(0.03); 
 \draw[fill=black] (-1.6,-0.5) circle(0.03); 
 \draw[fill=black] (-1.5,-0.9) circle(0.03);
\node[right] at (2.5,0) {$+$};
\end{tikzpicture}
\begin{tikzpicture}[line width=1 pt, scale=0.4, rotate=0,baseline=0]
\draw(-2,2)--(0,0); \draw(-2,-2)--(0,0)--(1,0); \draw (2,0) circle (1);
 \draw[fill=black] (-1.5,0.9) circle(0.03); 
 \draw[fill=black] (-1.6,0.5) circle(0.03); 
 \draw[fill=black] (-1.65,0.0) circle(0.03); 
 \draw[fill=black] (-1.6,-0.5) circle(0.03); 
 \draw[fill=black] (-1.5,-0.9) circle(0.03);
\end{tikzpicture}
\begin{tikzpicture}[line width=1 pt, scale=0.4, rotate=0,baseline=0]
\draw(-2,2)--(0,0);  \draw(-2,-2)--(0,0);
\begin{scope}[shift={(0,0)},scale=0.9]
	   \draw[fill=white,thin] (0,0) circle (0.3);
            \draw[thin] (45:0.3)--(225:0.3);
            \draw[thin] (-45:0.3)--(135:0.3);
	\end{scope}
\node[right] at (-4,0) {$=$};
 \draw[fill=black] (-1.5,0.9) circle(0.03); 
 \draw[fill=black] (-1.6,0.5) circle(0.03); 
 \draw[fill=black] (-1.65,0.0) circle(0.03); 
 \draw[fill=black] (-1.6,-0.5) circle(0.03); 
 \draw[fill=black] (-1.5,-0.9) circle(0.03);
\end{tikzpicture}\ \ .
\ee
Here dots stand for $N$ external lines and the  crossed 
 vertex in the r.h.s.  corresponds to the  interaction vertex  $\sim \chi^{N}$.
 Its  precise  coefficient  may be determined  as follows.
 The  explicit form of  the l.h.s. of (\ref{A.8}) is  
\be
\la{A.9}
-\frac{1}{2}\frac{1}{b^{2}}\,(2b)^{N+2}\,g(z,z)+\frac{1}{2}\frac{1}{b^{2}}\,(2b)^{N+1}\frac{1}{b^{2}}
(2b)^{3}\,\int \mathsf{d}^{2}z'  g(z,z')\, g(z',z') = -\C_{1}\,(2b)^{N},
\ee
where we used (\ref{A.7}) and took into account the $\frac{1}{2}$ symmetry 
factor due to the tadpoles. 
In the {\sf ZZ} formulation we have $\C_{1}^{\sf ZZ}=1$  (see \rf{2.14}) 
and  thus the r.h.s.   corresponds to  $(2b)^{N}\,\frac{\chi^{N}}{N!}$, 
i.e. $b^{2}$ times the interaction term in the {\sf ZZ} Lagrangian \rf{2.4}.
Thus the one-loop 
 tadpoles in the {\sf ZZ}  formulation give
  the same contribution as  the  extra couplings in \rf{2.16}  in the AdS  formulation.

\subsection{One-loop  correction to the two-point function}

The one-loop correction to the connected two-point function is the sum of the 
bubble diagram
\iffa\begin{center}
\begin{tikzpicture}[line width=1 pt, scale=0.6,baseline=-0.4cm]
\draw[fill=black] (-1.8,0) circle (0.08); \draw[fill=black] (1.8,0) circle (0.08); 
\draw (-1.8,0)--(-0.8,0); \draw (0,0) circle(0.8);
\draw (1.8,0)--(0.8,0);
\node[left] at (-1.8,0) {$z_{1}$}; \node[right] at (1.8,0) {$z_{2}$};
\end{tikzpicture}
\end{center}\fi
\begin{align}
D(z_{1}, z_{2}) &=\begin{tikzpicture}[line width=1 pt, scale=0.4,baseline=-0.1cm]
\draw[fill=black] (-1.8,0) circle (0.08); \draw[fill=black] (1.8,0) circle (0.08); 
\draw (-1.8,0)--(-0.8,0); \draw (0,0) circle(0.8);
\draw (1.8,0)--(0.8,0);
\node[left] at (-1.8,0) {$z_{1}$}; \node[right] at (1.8,0) {$z_{2}$};
\end{tikzpicture}=
 \int\mathsf{d}^{2}z'\,\mathsf{d}^{2}z''\,g(z_{1}, z')\,\big[g(z', z'')\big]^{2}\,g(z'', z_{2})\no \\
&=  \int\mathsf{d}^{2}z'\,g(z_{1}, z')\,\widetildeB(z', z_{2}),\qquad\qquad 
\widetildeB(z_{1},z_{2}) = \int\mathsf{d}^{2}z'\,\big[g(z_{1}, z')\big]^{2}\,g(z', z_{2}),
\la{A.11}
\end{align}
plus the tadpole diagrams contribution (\ref{A.8}) that, 
using \rf{A.7},\rf{A.9}, is given by 
  $\C_{1}\,\widehat \rB(z_{1},z_{2})$
with 
\be\la{a16}
\widehat \rB(z_{1},z_{2}) = \int\mathsf{d}^{2}z'\,g(z_{1},z')\,g(z',z_{2}).
\ee
Let us begin with the evaluation of $B(z_{1},z_{2})$.
The result should  be the   function of the $SU(1,1)$ invariant  $\eta(z,z')= \big|\frac{z-z'}{1-z\,\overline{z}'}\big|^{2}$ in \rf{2.9}. 
We may use the  3 parameters of  $SU(1,1)$  to first 
 set $z_{1}=0$ and $z_{2}=R\in\mathbb R$
and then   identify $R^{2}$  with $\eta$ in the final result.\footnote{One 
 can always find an $SU(1,1)$ transformation sending $z_{2}\to R>0$ and $z_{1}\to 0$: 
 the corresponding parameters are 
$b = -a\,z_{1}$,  $\ |a| = 1/(1-|z_{1}|^{2})$. 
From $R = (a z_{2}-a z_{1})/(-\overline a\, \overline z_{1}z_{2}+
\overline a) = \frac{a}{\overline a}\, (z_{2}-z_{1})/(1-z_{2}\overline z_{1})$ we find $R^{2} = \eta(z_{2},z_{1})$
and determine also  $\arg a$.}
Denoting  $g(z,0) \equiv  
g(|z|^{2})$
and $x=|z'|^{2}$, we get (see \rf{a3} for the expression for  $g_0$) 
\begin{align}
\widetildeB(0,R) &=  
\int\mathsf{d}^{2}z'\,\big[g(|z'|^{2})\big]^{2}\,g(z', R) =\int_{0}^{1}\frac{dx}{(1-x)^{2}}\,[g(x)]^{2}\,g_{0}(x, R^{2}) \notag \\
&= b_{0}(R^{2})\,\int_{0}^{R^{2}}\frac{dx}{(1-x)^{2}}\,[g(x)]^{2}\,a_{0}(x)
+a_{0}(R^{2})\,\int_{R^{2}}^{1}\frac{dx}{(1-x)^{2}}\,g(x)^{2}\,b_{0}(x)\notag \\
&= \frac{1}{8}-\frac{R^{2}\,\log^{2}(R^{2})}{8\,(R^{2}-1)^{2}}.\la{a13}
\end{align}
The  result  is thus  
\cite{Zamolodchikov:2001ah} 
\be
\la{A.13}
\widetildeB(z_{1},z_{2}) =  \frac{1}{8}-\frac{\eta\,\log^{2}\eta}{8\,(1-\eta)^{2}},\qquad \qquad \eta=\eta(z_{1},z_{2}).
\ee
It is then  straightforward to find the expression for $D(z_{1},z_{2})$ in \rf{A.11}. 
We  choose again 
$z_{1}=R>0$ and $z_{2}=0$ to get 
\begin{align}
D(z_1,z_2) & =  D(R,0) = \int\mathsf{d}^{2}z'\,g(R, z')\,\widetildeB(z', 0) = 
 \int\mathsf{d}^{2}z'\,g(R, z')\,\Big[ \frac{1}{8}-\frac{|z'|^{2}\,\log^{2}(|z'|^{2})}{8\,(1-|z'|^{2})^{2}}\Big]\notag \\
 &= \int_{0}^{1}\frac{dx}{(1-x)^{2}}\,g_{0}(R^{2}, x)\,
 \Big[ \frac{1}{8}-\frac{x\,\log^{2}x}{8\,(1-x)^{2}}\Big]\notag \\
 &= b_{0}(R^{2})\int_{0}^{R^{2}}\frac{dx}{(1-x)^{2}}\,a_{0}( x)\,
 \Big[ \frac{1}{8}-\frac{x\,\log^{2}x}{8\,(1-x)^{2}}\Big]+
 a_{0}(R^{2})\int_{R^{2}}^{1}\frac{dx}{(1-x)^{2}}\,b_{0}( x)\,
 \Big[ \frac{1}{8}-\frac{x\,\log^{2}x}{8\,(1-x)^{2}}\Big]\notag \\
 &= \frac{\eta ^2 \log ^2\eta }{64 (\eta -1)^2}-\frac{4 \eta +9}{96 \
(\eta -1)}-\frac{\eta  \log \eta }{48 (\eta -1)}+\log (1-\eta ) \
\Big[\frac{1}{48} -\frac{(\eta +1) \log \eta }{48 (\eta \
-1)}\Big]\notag \\
&\ \ \ -\frac{(\eta +1) \text{Li}_2(\eta )}{96 (\eta -1)}-\frac{39+\pi^{2}}{576}\frac{1+\eta}{1-\eta},
\la{a15}
\end{align}
where in the last step  we  replaced   $R^{2}$  by $\eta$. 
%
%
The tadpole contribution (\ref{a16}) is worked out as in 
\rf{a13},\rf{a15}. We get 
\begin{align}
\widehat \rB(R,0) &= \int\mathsf{d}^{2}z'\,g(R,z')\,g(z',0) =  \int\mathsf{d}^{2}z'\,g(R,z')\,g(|z'|^{2}) =
 \int_{0}^{1}\frac{dx}{(1-x)^{2}}\ g_{0}(R^{2},x)\,g(x) \notag \\
&= b_{0}(R^{2})\,\int_{0}^{R^{2}}\frac{dx}{(1-x)^{2}}\ a_{0}(x)\,g(x) 
+a_{0}(R^{2})\,\int_{R^{2}}^{1}\frac{dx}{(1-x)^{2}}\ b_{0}(x)\,g(x) \notag \\
&= -\frac{\eta  \log \eta }{6 (\eta -1)}+\log (1-\eta ) \
\Big[\frac{1}{6}-\frac{(\eta +1) \log \eta }{12 (\eta -1)}\Big]-\frac{(\eta \
+1) \text{Li}_2(1-\eta )}{6 (\eta -1)}-\frac{1}{6}.\la{a17}
\end{align}
Taking into account the combinatoric coefficients and the couplings \footnote{
For instance, the diagram associated with $D(z_{1},z_{2})$ must be multiplied
by $(-8)^{2}$ from the cubic couplings and by the symmetry factor $\frac{1}{2}$ giving $+32$
in (\ref{A.17}).
} 
the total one-loop correction to the  connected part of the two-point function is  (cf. (\ref{2.20}))
\begin{align}
\la{A.17}
32\,& D(z_{1},z_{2})-4\,\C_{1}\,\widehat{\rB}(z_{1},z_{2}) = 
\frac{3}{2}+\frac{\eta^{2}\log^{2}\eta}{2\,(1-\eta)^{2}}-\frac{1+\eta}{1-\eta}\,\text{Li}_{2}(1-\eta)
\notag \\
&+\frac{2}{3}\,(\C_{1}-1)\,\Big(
1-\frac{\eta\,\log\eta}{1-\eta}-\log(1-\eta)\,\Big[1+\frac{(1+\eta)\,\log\eta}{2\,(1-\eta)}\Big]
-\frac{1+\eta}{1-\eta}\,\text{Li}_{2}(1-\eta)
\Big).
\end{align}
For $\C_{1}=1$ this gives $\Sigma(z_{1},z_{2})$ in (\ref{2.21}) and 
agrees with the result of   \cite{Menotti:2004uq}.\foot{Note also that 
$\widehat{\rB}(z_{1},z_{2})$ coincides with
the expression in  Eq.(86) of \cite{Menotti:2004uq} because the r.h.s. of that equation
 is given by the 
 diagrams associated with $\widehat{\rB}(z_{1},z_{2})$.}

\section{On the one-point function in the {\sf ZZ} and {\sf AdS} formulations}
\la{app:tadpole}

Let us make  few comments on the   tadpole contributions in the \ZZ and AdS   formulations
 discussed in section 2.1.
As follows from   (\ref{2.18}), 
the tadpole $\langle \chi(z)\rangle^{\sf ZZ}$ has a non-constant
 $\mc O(b)$ contribution that combined with the classical background \rf{2.3} 
 changes its coefficient from $b^{-1}$ to 
 $Q=b^{-1} + b$. The  constant part of the $\mc O(b)$  term  in \rf{21.7},\rf{2.18}
  is  regularization dependent. 
  
  In general, the  two-loop purely tadpole  corrections to the one-point function are  given by 
\be
\mc T_2(z)= \ \ \begin{tikzpicture}[line width=1 pt, scale=0.6,baseline=-0.1cm]
\draw (180:1)--(0,0)--(30:1); \draw (0,0)--(-30:1);
\draw (30:1.5) circle(0.5);\draw (-30:1.5) circle(0.5);
\draw[fill=black] (180:1) circle (0.1); 
\node[above] at (180:1) {$z$};
\end{tikzpicture}\ \ +
\begin{tikzpicture}[line width=1 pt, scale=0.6,baseline=-0.1cm]
\draw (180:1)--(0,0); \draw(90:0.5) circle(0.5); \draw(0,0)--(1,0); \draw(1.5,0) circle(0.5);
\draw[fill=black] (180:1) circle (0.1); 
\node[above] at (180:1) {$z$};
\end{tikzpicture}\ \ +
\begin{tikzpicture}[line width=1 pt, scale=0.6,baseline=-0.1cm]
\draw (180:1)--(1,0); \draw(1,0.5) circle(0.5);\draw(1,-0.5) circle(0.5);
\draw[fill=black] (180:1) circle (0.1); 
\node[above] at (180:1) {$z$};
\end{tikzpicture}\ \ . \la{b1}
\ee
These are absent   in the AdS formulation (where all tadpoles vanish, see \rf{2.15})
but are   non-zero  and  {\it constant}  in the {\sf ZZ}   formulation. 
While   separate  contributions to \rf{b1} are not 
 $SU(1,1)$ covariant, their sum is  and it may be written  as (see also \cite{menotti2006liouville})
\be
\la{B.2}
\mc T_2(z) \sim \int \mathsf{d}^{2}z' g(z,z')\,\bigg[
\begin{tikzpicture}[line width=1 pt, scale=0.6,baseline=-0.1cm]
\draw (180:1)--(0,0); \draw(0.5,0) circle(0.5);
\draw[fill=black] (180:1) circle (0.1); 
\node[above] at (180:1) {$z'$};
\end{tikzpicture}\ \ +
\begin{tikzpicture}[line width=1 pt, scale=0.6,baseline=-0.1cm]
\draw(180:0.5) circle(0.5);
\draw[fill=black] (180:1) circle (0.1); 
\node[left] at (180:1) {$z'$};
\end{tikzpicture}\ \ 
\bigg]^{2} \sim \int \mathsf{d}^{2}z' \, g(z,z')\ . 
\ee
This is just a $z$-independent constant  (as one can see, e.g.,  by 
using   $SU(1,1)$ invariance  or  comparing  (\ref{A.4}) and   (\ref{A.6})). 
In a general regularization scheme  in \rf{2.13}   one thus  finds that 
 $\mc T_2(z)\sim \C_{1}^{2}$  (cf. \rf{A.7}). 
 This is consistent  with the vanishing of $\mc T_2(z)$ in the 
Hadamard-like regularization  (where  $\C_{1}=0$, $\C_{2}=\text{const}$ in  \rf{2.13}) discussed in \cite{Menotti:2004uq}.

\def \CCC  {{\rm C}}

\section{Tree level Witten diagram  contributions  to  $\langle\Phi(t_{1})\cdots \Phi(t_{4})\rangle$}
\la{app:tree}

At the tree level, the  connected  four-point boundary correlator 
$\CCC(t_1, ...,t_4) \equiv \langle \Phi(t_{1})\cdots \Phi(t_{4})\rangle_{\rm conn}$
is given by the sum of the contact and
the  exchange diagrams  (cf. $C_{4,0}^{\rm cont}$
and $C_{4,0}^{\rm exch}$ in (\ref{4.4})).
Their separate  contributions is   straightforward to 
 compute   \cite{Ouyang:2019xdd}:
\begin{align}
\CCC_{4,0}^{\rm cont}(t_{1},\dots, t_{4})\,b^{2} &=    \ \ 
\begin{tikzpicture}[line width=1 pt, scale=0.4, rotate=0,baseline=-0.1cm]
\coordinate (A1) at (135:2);  \coordinate (A2) at (45:2);  
\coordinate (A3) at (-45:2);   \coordinate (A4) at (-135:2);
\draw[densely dashed] (0,0) circle (2);
\draw (A1)--(0,0)--(A2); \draw (A3)--(0,0)--(A4);
\draw[fill=black] (A1) circle (0.1); \draw[fill=black] (A2) circle (0.1); 
\draw[fill=black] (A3) circle (0.1); \draw[fill=black] (A4) circle (0.1); 
\end{tikzpicture}\ \  = -\frac{1024}{81\pi}\,b^{2}\,D_{2,2,2,2}  \\
\CCC_{4,0}^{\rm exch}(t_{1}, \dots, t_{4}) \, b^2 &=  \ \ 
\begin{tikzpicture}[line width=1 pt, scale=0.4, rotate=0,baseline=-0.1cm]
\coordinate (A1) at (135:2);  \coordinate (A2) at (45:2);  
\coordinate (A3) at (-45:2);   \coordinate (A4) at (-135:2);
\coordinate (B1) at (-1,0); \coordinate (B2) at (1,0);
\draw[densely dashed] (0,0) circle (2);
\draw (A1)--(B1)--(A4); \draw (A3)--(B2)--(A2); \draw (B1)--(B2);
\draw[fill=black] (A1) circle (0.1); \draw[fill=black] (A2) circle (0.1); 
\draw[fill=black] (A3) circle (0.1); \draw[fill=black] (A4) circle (0.1); 
\end{tikzpicture}\ \ +\text{crossed} = \frac{512}{81\pi}\,b^{2}\,\big(
t_{12}^{-2}\,D_{1,1,2,2}+t_{13}^{-2}\,D_{1,2,1,2}+t_{14}^{-2}\,D_{1,2,2,1}\big), \no
 \end{align}
where the function 
 $D_{n,k,l,m}(t_{1}, \dots, t_{4})$
 are the standard AdS integrals 
  defined in \cite{DHoker:1999kzh,Dolan:2000ut,Dolan:2003hv}. 
Their explicit evaluation gives
\begin{align}
\la{D.2}
\CCC_{4,0}^{\rm cont} &= \frac{64}{81\,t_{12}^{4}\,t_{34}^{4}}\,\bigg[
\frac{2 (\chi ^2-\chi +1) \chi ^2}{(1-\chi )^2}+(2 
\chi ^2+\chi +2) \chi  \log (1-\chi )+\frac{(2 \chi ^2-5 \chi +5) 
\chi ^4 \log \chi }{(1-\chi )^3}
\bigg],\notag \\
\CCC_{4,0}^{\rm exch} &= \frac{64}{81\,t_{12}^{4}\,t_{34}^{4}}\,\bigg[
\frac{(\chi ^2-\chi +1) \chi ^2}{(1-\chi )^2}-(2 \chi ^2+\chi +2) 
\chi  \log (1-\chi )-\frac{(2 \chi ^2-5 \chi +5) \chi ^4 \log \chi}{(1-\chi )^3}\bigg]\ , 
\end{align}
where  $\chi=\frac{t_{12}t_{34}}{t_{13}t_{24}}$ is the 1d cross-ratio.
 Note that  these two   contributions have  non-trivial  logarithmic 
dependence on the boundary points $t_{1}, \dots, t_{4}$.  However, 
all the logarithms cancel in their sum    which becomes 
simply proportional to the  conformal 
factor $\mc K_{4}$ in (\ref{4.9}):
\begin{align}
& \CCC_{4,0}^{\rm cont}(t_{1},\dots, t_{4}) +
\CCC_{4,0}^{\rm exch}(t_{1},\dots, t_{4})  \notag \\
&\quad = \frac{64}{27}\,\frac{1}{t_{12}^{4}\,t_{34}^{4}}\,
\frac{\chi^{2}\,(1-\chi+\chi^{2})}{(1-\chi)^{2}} = \frac{32}{27}\,\bigg(
\frac{1}{t_{12}^{2}\,
t_{23}^{2}\,t_{34}^{2}\,t_{14}^{2}}
+\frac{1}{t_{13}^{2}\,t_{24}^{2}\,t_{14}^{2}\,t_{23}^{2}}
+\frac{1}{t_{12}^{2}\,t_{24}^{2}\,t_{34}^{2}\,t_{13}^{2}}
\bigg),\la{cc3}
\end{align}
reproducing the value  for the coefficient  $C_{4,0}$ in  (\ref{4.4}).

 Note that this   cancellation  does not happen if the two contributions are combined with different weights  like in  (\ref{4.6})  where $C_{4,0}^{\rm cont}$  and $C_{4,0}^{\rm exch}$ 
 should  then  be understood as   the functions in \rf{D.2} divided by the factor $\mc K_{4}$. 
 

\section{Global conformal block decomposition of the four-point  correlator }
\la{app:tttt}

The four-point correlator in \rf{1.11}--\rf{1.13}  has a simple  dependence on the boundary 
points
without any logarithmic  terms of the 1d  cross ratio $\chi$.
This  means that the 1d  conformal operators  appearing in the OPE 
of the   boundary   correlator \rf{1.11}  will have no anomalous dimensions
(cf. \cite{Giombi:2017cqn,Beccaria:2019dws}).
The same    structure  is fixed by  the Virasoro symmetry in the case of the 
stress  tensor correlator \rf{1.5}. 
In view of this connection it is of  some 
 interest to discuss  the  equivalent OPE decomposition of 
the  stress  tensor correlator in terms of the  conformal blocks of the global 
part of the 2d   conformal symmetry (which is effectively equivalent to 1d conformal block decomposition). 

\iffa 
The four-point function of the stress tensor in (\ref{1.5}) may be analyzed in terms of the global conformal data
of operators scattered in a certain kinematical channel. This kind of analysis in boundary one dimensional CFTs
has been quite useful in the recent studies  of the defect CFT associated with Wilson loops in $\mc N=4$ SYM
in 4d \cite{Giombi:2017cqn,Beccaria:2019dws}. Notice however that (\ref{1.5}) does not involve logarithms of 
the cross-ratio implying no corrections to the classical dimensions of the exchanged operators. 
\fi

One  can   represent the correlator in \rf{1.5} as 
\begin{align}
\la{C.1}
&\langle T(z_{1})\, T(z_{2})\,T(z_{3}) \,T(z_{4})\rangle = \frac{1}{z_{12}^{4}\, z_{34}^{2}}\,G(\chi),\qquad\qquad 
\chi \equiv  \frac{z_{12}z_{34}}{z_{13}z_{24}},\\
\la{C.2}
&G(\chi) = \frac{c^{2}}{4}\,\Big[1+\chi^{4}+\frac{\chi^{4}}{(1-\chi)^{4}}\Big]
+2\,c\,\frac{\chi^2(1-\chi+\chi^{2})}{(1-\chi)^{2}} \ , 
\end{align}
where the  $\frac{c^{2}}{4}$ term is the "generalized free field" part. 
From  the global $SL(2,\mathbb R)$ invariance  we should  have 
\be
\la{C.3}
G(\chi) = \sum_{n=0}^{\infty} c_{n}\,\mathsf{F}_{n}(\chi), \qquad 
\mathsf{F}_{n}(\chi) = \chi^{n}\,_{2}F_{1}(n,n,2n; \chi),\qquad 
c_{n} = \sum_{\mc O_{n}}\frac{\CC_{3, \mc O_{n}}^{2}}{\CC_{2, \mc O_{n}}},
\ee
where  $\{\mc O_{n}\}$  stand for the dimension $n$ 
 quasi-primary fields (with $T$  corresponding to $\{\mc O_{2}\}$) 
   in the vacuum module 
and 
$\CC_{3, \mc O}$ and $\CC_{2, \mc O}$ are the coefficients in 
$\langle TT\mc O\rangle$ and $\langle \mc O \mc O\rangle$.
One   can show 
by a direct analysis of the algebraic
expression in (\ref{C.2})  that  \cite{Osborn:2012vt}
\be
c_{2p+2} = \Big[\frac{c^{2}}{144}\,(2p-1)_{6}+2\,c\,\big[1+2p(2p+3)\big]\Big]\frac{(2p)!(2p+1)!}{(4p+1)!}, \qquad c_{2p+1}=0, \ \ \ \  p=0, 1, 2, \dots \ .  
\ee
Thus 
\begin{align}
\la{C.5}
G(\chi) &= \frac{c^2}{4}+2\,c\,\mathsf{F}_{2}(\chi)+\Big(\frac{c^{2}}{2}+\frac{11}{5}\,c\Big)
\,\mathsf{F}_{4}(\chi)+\Big(\frac{10}{9}\,c^{2}+\frac{29}{63}\,c\Big)\,\mathsf{F}_{6}(\chi)
+\mc O(\chi^{8}) \ . 
\end{align}
Here 
$c_{0}=\frac{c^{2}}{4}$ and $c_{2}=2c$ are associated with $\mc O_{0}=\mathbb{I}$
and $\mc O_{2}=T$.
At level 4 we have  the quasi-primary  field (appearing in the
regular part of the OPE  of  $T(z)T(z')$)
\be
\Lambda_{4} = (TT)-\tfrac{3}{10}\,T'',
\ee
where  the brackets denote the normal ordering. Using that 
$
\CC_{2, \Lambda_{4}} = \CC_{3,\Lambda_{4}}=\frac{c(22+5c)}{10},
$
we confirm the value of $c_{4}$ in \rf{C.5}. 
At level 6 one finds   two orthogonal quasi-primaries
\begin{align}
\Lambda_{6}^{(a)} &= (T(TT))-\tfrac{9}{8} (T'\,T')-\tfrac{1}{112}\,T'''',  \\
\Lambda_{6}^{(b)} &= (T(TT))
-\tfrac{28 c^2+1050 c+2735 }{18 (42 c+67)}\,(T''\,T)+\tfrac{35 c^2+462 c+2062}{18 (42 c+67)}
(T'\,T')-\tfrac{5 c^2+228 c+553}{108 (42 c+67)}\,T'''',\no 
\end{align}
and a straightforward calculation gives
\begin{align}
\CC_{2, \Lambda_{6}^{(a)}} &= \tfrac{3}{112} c (28 c^2+1050 c+2735), & 
\CC_{3,\Lambda_{6}^{(a)}} &= \tfrac{3}{28} c (42 c+67),  \\
\CC_{2, \Lambda_{6}^{(b)}} &= \tfrac{c (2 c-1) (5 c+22) (7 c+68) (28 c^2+1050 c+2735)}{36 (42 c+67)^2}, & 
\CC_{3,\Lambda_{6}^{(b)}} &= -\tfrac{c (2 c-1) (5 c+22) (7 c+68)}{9 (42 c+67)}.\no
\end{align}
As a result, $c_{6}$ in \rf{C.3}  is given by the sum 
\be
c_{6} = \tfrac{3 c (42 c+67)^2}{7 (28 c^2+1050 c+2735)}+\tfrac{4 c (2 c-1) (5 c+22) (7 c+68)}{9 (28 c^2+1050 c+2735)}
= \tfrac{10}{9}\,c^{2}+\tfrac{29}{63}\,c,
\ee
in agreement with (\ref{C.5}).

\bibliography{BT-Biblio}
\bibliographystyle{JHEP}

\end{document}